\documentclass[preprintnumbers,prd,tightenlines,nofootinbib,superscriptaddress]{revtex4}

\usepackage{amsfonts,amssymb,amsthm,bbm}
\usepackage{amsmath}

\usepackage{color,psfrag}
\usepackage[dvips]{graphicx}

\usepackage{subcaption}
\usepackage{subcaption}
\usepackage{graphicx}
\usepackage{xcolor}
\usepackage{tabularray}
\usepackage{physics}
\usepackage{csquotes}
\usepackage{booktabs}

\usepackage{hyperref}  

\newcommand{\cL}{{\mathcal L}}
\newcommand{\cH}{{\mathcal H}}

\newcommand{\be}{\begin{equation}}
\newcommand{\ee}{\end{equation}}
\newcommand{\beq}{\begin{eqnarray}}
\newcommand{\eeq}{\end{eqnarray}}
\newcommand{\bes}{\begin{eqnarray}}
\newcommand{\ees}{\end{eqnarray}}

\newcommand{\lp}{\left(}
\newcommand{\rp}{ \right)}
\newcommand{\lc}{\left[}
\newcommand{\rc}{\right]}

\newcommand{\f}{\frac}

\def\nn{\nonumber}

\def\om{\omega}

\def\ba{\tilde{a}}

\begin{document}

\preprint{YITP-24-58, RIKEN-iTHEMS-Report-24}

\title{Scalar Quasi-Normal Modes of a Loop Quantum Black Hole}

\author{{\bf Etera R. Livine}}\email{etera.livine@ens-lyon.fr}
\affiliation{ENSL, CNRS, Laboratoire de Physique, F-69342 Lyon, France}

\author{{\bf Clara Montagnon}}\email{clara.montagnon@ens-lyon.fr}
\affiliation{ENSL, CNRS, Laboratoire de Physique, F-69342 Lyon, France}

\author{{\bf Naritaka Oshita}}\email{naritaka.oshita@yukawa.kyoto-u.ac.jp}
\affiliation{Center for Gravitational Physics and Quantum Information,
Yukawa Institute for Theoretical Physics, Kyoto University, 606-8502, Kyoto, Japan}
\affiliation{The Hakubi Center for Advanced Research, Kyoto University,
Yoshida Ushinomiyacho, Sakyo-ku, Kyoto 606-8501, Japan}
\affiliation{RIKEN iTHEMS, Wako, Saitama, 351-0198, Japan}

\author{{\bf Hugo Roussille}}\email{hugo.roussille@ens-lyon.fr}
\affiliation{ENSL, CNRS, Laboratoire de Physique, F-69342 Lyon, France}

\begin{abstract}
We compute the Quasi-Normal Mode (QNM) frequencies for scalar perturbations for modified Schwarzschild black holes  in Loop Quantum Gravity. We study the singularity-free polymerized metric characterised by two parameters encoding loop quantum effects: the minimal area gap $a_0$ and the polymeric deformation parameter $P$.
We perform numerical computations using Leaver's continued fraction method and compare our results to other semi-analytical methods and existing literature. We study the effects on the QNM spectrum of variation of both deformation parameters and systematically compare to the standard Schwarzschild case. In particular we find that the scalar fundamental mode is modified from the third decimal for values of $P$  in accordance with the most recent astrophysical constraints. We also show that qualitative differences arise for highly damped modes: on the one hand, a new crossing of the imaginary axis occurs for high values of $a_0$ and, on the other hand, increasing $P$ produces a positive shift of the real part and an increase of the spacing in imaginary part between modes.
\end{abstract}

\maketitle

\makeatletter
\def\l@subsection#1#2{}
\def\l@subsubsection#1#2{}
\makeatother
\tableofcontents
\newpage

\section*{Introduction}

General Relativity (GR) is a very successful theory of gravity, which correctly explained and predicted experimental results in many different energy regimes, from the cosmological observations of the PLANCK collaboration~\cite{Planck:2018nkj} to the dynamics of the Hulse-Taylor binary pulsar system~\cite{Hulse:1974eb, Weisberg:1981bh}. However, until a few years ago, direct testing of the dynamics of black holes (BHs) -- the most compact objects predicted by GR, exhibiting the highest gravitational fields -- had not yet been performed. The recent direct detection of gravitational waves (GWs) emitted by a binary BH merger~\cite{LIGOScientific:2016aoc} made testing GR in this strong field regime possible. While present data obtained by the LIGO-Virgo-KAGRA collaboration is compatible with GR predictions up to the precision of measurement, the increasing number of events and the expected construction of higher-precision detectors, such as LISA, Einstein Telescope, or Cosmic Explorer, will  either lead to a larger precision validation of GR or emphasise the existence of deviations from the standard theory.

Testing GR through the study of GWs emitted by BH mergers requires to model the GW waveforms theoretically and compare it to the measurements. Three regimes are distinguished in the GW waveform: the inspiral, the merger and the ringdown. The inspiral phase corresponds to the beginning of the merger when the two BHs orbit each other and lose energy via the emission of GWs. Obtaining the waveform for this phase is a complex task since it requires finding a solution to Einstein's equations describing non-linear superpositions of BHs. The merger phase describes the merging of the two event horizons and is also highly non-linear. In the last part of the signal, the ringdown, GWs are emitted by the resulting single out-of-equilibrium BH  settling down towards a stationary state. This part is the easiest to model, since the system can a priori be described by linear perturbations around a stationary BH.

The response of a BH to linear perturbations is composed by an initial GW burst, followed by a sum of damped sinusoids and ending with a power-law tail. 
The complex frequencies building the sum of damped sinusoids are called Quasi-Normal Modes (QNMs) \cite{Kokkotas:1999bd, Nollert:1999ji, Berti:2009kk, Konoplya:2011qq, franchini2023testing}. Their measurement via what is now called {\it black hole spectroscopy} is an invaluable tool to test GR since it probes the perturbative dynamics of the theory as well as the stationary background solution~\cite{Berti:2005ys, Berti:2018vdi, Maselli:2019mjd}. Until now, BH spectroscopy led to validation of GR predictions up to the precision of measurements~\cite{LIGOScientific:2016lio, Ghosh:2021mrv}.
%

Despite the numerous experimental successes of GR, the theory is failing to provide a good description of the fate of BHs \cite{Penrose:1964wq} and of the very early universe \cite{Hawking:1970zqf}, amongst others. These physical phenomena highlight the need for a theory reconciling quantum mechanics and gravity. Loop Quantum Gravity (LQG) is a non-perturbative and background independent approach to quantizing general relativity, and is now one promising candidate for a theory of Quantum Gravity. It is nevertheless still very challenging to extract predictions directly from the full theory. Indeed although it provides a very precise description of the quantum space-time directly at the Planck scale, one needs to coarse-grain it by several order of magnitude to get rigorous predictions at astrophysical scales. Determining the resulting phase diagrams is still a very active line of research and, as a consequence, the use of effective models has become common practice.
In particular, concepts, tools and methods, from LQG can be used to construct effective LQG BH models expressed as quantum corrected Schwarzschild solutions. The usual procedure consists in
regularizing GR's Hamiltonian constraint, with holonomy and volume corrections based on the polymerisation technique at the heart of LQG.
%
%
This is why effective LQG BHs are also referred as polymerized BHs.
One of the first proposal was provided by Modesto in 2008 \cite{Modesto:2008im}.
Based on a constant polymeric deformation parameter $\delta$, this work showed that the singularity was replaced by a bounce of the 2-sphere (at given time and radial coordinates) on the minimal area $a_0$ of LQG. 
%
%
Since then, several effective LQG BH solutions have been constructed in the past years \cite{Modesto:2008im,Peltola_2009a,Peltola_2009b,Gambini_2013,Gambini_2014,Gambini_2020,Bodendorfer_2019,Bodendorfer_2021a,Bodendorfer_2021b,Ashtekar_2018,Ashtekar_2020,Kelly_2020a,Kelly_2020b,Alonso_Bardaji_2022,Ongole_2024}. Their common feature, besides being based on LQG methods, is that they all lead to a resolution of the BH space-time singularity. 

The lack of discriminating test for these effective quantum BH ansatz currently prevents us from sorting them out. Even though the current status of experimental QNM detection is not sufficiently developped as for now, one possibility is to theoretically compute the QNM of these effective BH models on order to compare them with hypothetical future detection of deviations from GR predictions. 

In this paper, we compute the QNMs of Modesto's effective BH \cite{Modesto:2008im}, in order to investigate the effects of the quantum parameters $a_0$ and $P(\delta)$ on the QNM spectrum and to characterise the deviations from the Schwarzschild spectrum. The QNMs of this polymerized BH have already been examined \cite{Moulin:2019ekf,{Liu:2020ola,Momennia:2022tug,Yang:2023gas}}. We stand out from these analyses by considering wide ranges of values for both the quantum parameters $a_0$ and $P$ and by pushing the study to highly damped QNMs instead of merely investigating the first few modes. We focus here on scalar QNMs, for the sake of avoiding ambiguities arising in the effective propagation of electromagnetic and gravitational fields in the infra-red regime of LQG.

Various semi-analytical and numerical methods have been developed for QNM computation. One of the most efficient techniques has been proposed by Leaver in 1985 \cite{Leaver:1985ax} and relies on a reformulation of the eigenvalue problem into a question of finding the roots for a continued fraction.
A key point is to impose the appropriate physically motivated boundary conditions: the modes should be ingoing at the BH horizon and outgoing at spatial infinity. 
Leaver's approach, often referred to as the \enquote{continued fraction method}, allows for a high-accuracy computation of a large number of QNMs~\cite{Onozawa:1996ux, Berti:2003jh, Berti:2004um, Leaver:1990zz, Berti:2005eb, Yoshida:2003zz}.
In this article, we apply this method to the polymerized BH of Modesto \cite{Modesto:2008im} and describe the changes induced in the QNM spectrum.
We compare our results with subcases already considered in the literature and cross-check our numerical computations using other existing QNM computation methods, such as the WKB method~\cite{Schutz:1985km, Iyer:1986np, Iyer:1986nq, Konoplya:2003ii,Matyjasek:2017psv} (see~\cite{Konoplya:2019hlu} for a review) and the spectral decomposition method~\cite{Jansen:2017oag}.

This paper is structured as follows. In the first section, we review the formalism for the study of scalar field perturbations around a Schwarzschild BH and we describe the continued fraction method. We explain in detail how one can solve the continued fraction equation in order to compute QNMs with great accuracy, even for the high overtones. In section~\ref{sec:LQG-BH}, we present the polymerized BH, and we show how to adapt the continued fraction method to compute QNMs for this background metric. Finally, we present our numerical results and discuss  their interpretation in section~\ref{sec:results}. We conclude with perspectives and outlook.
We also provide additional details about our computations and comparisons with other existing numerical methods in appendices.

\section{Scalar perturbations of a Schwarzschild black hole}
\label{sec:pert-Schwa}

We start by reviewing the numerical computation of QNMs via Leaver's continued fraction method in the case of a Schwarzschild BH. After obtaining the well-known master equation for scalar perturbations on the Schwarzschild metric, we will sum up how QNMs are computed.

\subsection{Master equation}

The Schwarzschild BH metric is given by
\begin{align}
    &\dd{s}^2 = -f(r)\dd{t}^2 + \frac{\dd{r}^2}{g(r)} + h(r) \dd{\Omega}^2\,,\label{metric}\\
    &f(r) = g(r) = 1 - \frac{r_s}{r} \,,\quad h(r) = r^2 \,,
    \label{schwa}
\end{align}
where $r_s$ is twice the ADM mass of the BH and $\dd{\Omega}^2 = \dd{\theta}^2 + \sin^2{\theta}\dd{\varphi}^2$ is the line element on the 2-sphere. Let us now consider a test scalar field $\Phi$ propagating on this background. It satisfies the Klein-Gordon equation: 
\begin{equation}
    \nabla_{\mu}\nabla^{\mu}\Phi = \frac{1}{\sqrt{-g}}\partial_{\mu}(\sqrt{-g}g^{\mu\nu}\partial_{\nu}\Phi) =0 \,,
    \label{KG}
\end{equation}
where $g$ refers to the metric determinant and $g^{\mu\nu}$ to the inverse metric.
Since the background spacetime is static and spherically symmetric, we use the following decomposition for the field $\Phi$:
\begin{equation}
    \Phi(t,r,\theta,\varphi)= \sum_{l,\,m} \frac{\Psi_{l m}(t, r)}{\sqrt{h(r)}}Y_{l m}(\theta,\varphi),
\end{equation}
where $Y_{l m}$ represents a spherical harmonic. At first order, perturbations with different values of $l$ or $m$ do not couple due to the spherical symmetry of the spacetime. To make notations as simple as possible, we thus drop the $(l, m)$ indices. Furthermore, we introduce the Fourier transform in time $\hat{\Psi}$ of the scalar field,
\begin{equation}
    \Psi(t, r) = \frac{1}{\sqrt{2\pi}} \int\dd{\omega} \hat{\Psi}(\omega, r) e^{-i\omega t} \,.
\end{equation}
In the rest of this paper, we will work exclusively with $\hat{\Psi}(\omega, r)$.
We will keep the frequency variable $\om$ implicit and refer to this function simply as $\Psi(r)$.
%
The Klein-Gordon equation~\eqref{KG} then reduces to
\begin{equation}
    \dv[2]{\Psi}{x} + \qty[\omega^2 - V_\mathrm{sch}(r)]\Psi = 0,
    \label{schroSchwa}
\end{equation}
with the effective potential
\be
V_\mathrm{sch}(r) = \frac{r-r_s}{r} \qty(\frac{l(l+1)}{r^2} + \frac{r_s}{r^3}) \,.
\label{potSchwa}
\ee
We used the tortoise coordinate $x$ defined by 
\begin{equation}
    \dv{x}{r} = \frac{1}{\sqrt{f(r)g(r)}} \,,
    \label{eq:tortoise}
\end{equation}
which implies, in the Schwarzschild case, $x(r) = r + r_s\log(r-r_s)$ up to a constant. This coordinate is defined on the BH exterior $r\ge r_s$ and ranges from $-\infty$ to $+\infty$. The equation for $\Psi$ describes the propagation of waves at speed of light on the background, with scattering by an effective potential $V_\mathrm{sch}$. It is sometimes called the {\it master equation} for scalar perturbations of Schwarzschild BHs, and can be generalised to other spins of perturbations~\cite{Arbey:2021jif}.

\subsection{Continued fraction method}
\label{sec:contfrac}

The effective potential $V_\mathrm{sch}$ vanishes both at the horizon $r = r_s$ and at infinity.
This means that the reduced Klein-Gordon equation~\eqref{schroSchwa} has asymptotic plane-wave solutions at $x\sim\pm\infty$.
%
%
QNM solutions correspond to imposing that these waves be incoming at the horizon and outgoing at infinity:
\begin{subequations}
    \begin{align}
        &\Psi \sim e^{i\omega x} \sim e^{i\omega r}r^{i\omega r_s} \,, &\mkern-120mu(x\to+\infty,\,r \to +\infty)\\
        &\Psi \sim e^{-i\omega x} \sim (r-r_s)^{-i\omega r_s}  \,. &\mkern-120mu(x\to-\infty,\,r \to r_s)
    \end{align}
    \label{eq:asymp-schwa}
\end{subequations}
These boundary conditions, along with~\eqref{schroSchwa}, define an eigenvalue problem for the differential operator appearing in~\eqref{schroSchwa}, the eigenvalue being the frequency $\omega$. This problem is not self-adjoint. Therefore, in general, $\omega$ will be a complex number\footnotemark.
\footnotetext{However, for the perturbations to be stable, one expects the differential operator to be essentially self-adjoint, meaning that it possesses a unique self-adjoint extension. More information about the properties of the Schr\"odinger differential operator and applications to BH perturbation theory can be found in~\cite{Lewin2022, Horowitz:1995gi, Fabris:2020kog}.}

In general, there is no analytical solution\footnotemark{} for $\omega$. However, many semi-analytical and numerical methods have been developed to approximate the quasi-normal frequencies $\omega$. In the present paper, we use Leaver's method \cite{Leaver:1985ax}, also referred as the continued fraction method or the Frobenius method. This approach reformulates the eigenvalue problem into a continued fraction equation, the quasi-normal frequencies being its roots. It is one of the most efficient and accurate method to compute QNM~\cite{franchini2023testing}.
\footnotetext{One case where an analytical computation is possible is when the potential is the P\"oschl-Teller potential \cite{Ferrari:1984zz}.}

The starting idea is to formulate an ansatz for the radial solution of equation \eqref{schroSchwa} as a power series satisfying the QNM boundary conditions \eqref{eq:asymp-schwa}. The QNM domain of definition being $(r_s,+\infty)$, a power series of $r$ is not appropriate. The mapping
\be 
r \mapsto \frac{r-r_s}{r}
\ee
allows to shift the domain to $(0,1)$ and to obtain a well-defined ansatz. Taking into account the QNM boundary conditions \eqref{eq:asymp-schwa}, we get an ansatz for the radial wave function solution of the master equation \eqref{schroSchwa}:
\begin{equation}
    \Psi = e^{i\omega r}r^{i\omega r_s} \qty(\frac{r-r_s}{r})^{-i\omega r_s}  \sum_{n=0}^{+\infty} a_n \qty(\frac{r-r_s}{r})^n \,.
    \label{eq:ansatz-schwarzschild}
\end{equation}
Inserting this ansatz into the master equation~\eqref{schroSchwa} yields the following recurrence relation:
\begin{align}
    \alpha(n) a_{n+1} + \beta(n) a_n + \gamma(n) a_{n-1} =0\,, \qquad\qquad (n \geq 1)
    \label{eq:rec-rel-schwa}
\end{align}
along with the initial condition $\alpha(0) a_1 + \beta(0) a_0 = 0$.
In the remainder of the paper, we will set $r_s=1$ unless stated otherwise.
The sequences $\alpha(n)$, $\beta(n)$ and $\gamma(n)$ are given by
\begin{subequations}
\begin{align}
    \alpha(n) &= n^2+(2-2i\omega)n-2i\omega +1 \,,\nonumber\\
    \beta(n) &= -[2n^2+(2-8i\omega)n-8\omega^2-4i\omega+\lambda-1]\,,\nonumber\\
    \gamma(n) &= n^2-4i\omega n -4\omega^2\,.
\end{align}
\label{coef_schwa}
\end{subequations}
The computation of Leaver~\cite{Leaver:1985ax} then relies on the fact that the only solution to the three terms recurrence relation~\eqref{eq:rec-rel-schwa} leading to convergence of the power series in \eqref{eq:ansatz-schwarzschild} is the one such that the following continued fraction holds:
\begin{equation}
    \beta(0)-\frac{\alpha(0)\gamma(1)}{\beta(1)-}\frac{\alpha(1)\gamma(2)}{\beta(2)-}\frac{\alpha(2)\gamma(3)}{\beta(3)-}...=0
    \,,
    \label{eq:contfrac-schwa}
\end{equation}
where we use the standard condensed notation for the continued fraction.
%
This defines an equation for $\omega$ and solving it will yield the scalar QNMs for the Schwarzschild BH. 
In practice, this relation contains an infinite number of terms. 
%
%
In order to handle it numerically,
one truncates the continued fraction at some threshold $N$. As we will see, the value of $N$ strongly impacts the precision of the computed QNMs, and it is important to check that the value of $N$ used in the numerical method leads to precise stable results for the QNM numerics.

This method has been used in order to compute QNMs of the Schwarzschild BH and has then been extensively extended to other applications~\cite{Onozawa:1996ux, Berti:2003jh, Berti:2004um, Leaver:1990zz, Berti:2005eb, Yoshida:2003zz}. The present paper is dedicated to its application to the polymerized BH ansatz encoding LQG corrections to GR.

\section{Effective Loop Quantum Black Hole}
\label{sec:LQG-BH}

One of the first Schwarzshild BH solution with LQG-motivated corrections  was presented by Modesto in 2008~\cite{Modesto:2008im}. He considered a large class of Hamiltonian constraints expressed in terms of holonomy variables and for which the deviation from the classical Hamiltonian constraint is characterised by the polymeric parameter $\delta$. He selected the constraints compatible with spherical symmetry plus homogeneity, and obtained as a result that the Kretschmann scalar is regular in all space-time. This means that the 2-sphere (at given time and radial coordinates) bounces on a minimum radius. This minimal area corresponds to the {\it area gap} $a_0$ of the LQG theory. Indeed, LQG predicts a discrete spectrum for the area operator and $a_0$ gives its smallest non-vanishing eigenvalue, or in other words, {the smallest quantum of area}. In this scenario, this bounce, which naturally resolves the BH singularity, is thus a direct consequence of the quantization of geometric observables.

\subsection{Geometry}

The line element of the stationary and spherically symmetric LQG BH described in~\cite{Modesto:2008im} is given by the metric~\eqref{metric}, with the choice of metric functions being
\begin{equation}
    f(r) = \frac{(r-r_+)(r-r_-)}{r^4+a_0^2}(r+r_0)^2 \,,\qquad g(r) = \frac{(r-r_+)(r-r_-)}{r^4+a_0^2}\frac{r^4}{(r+r_0)^2}\,,\qquad h(r) = r^2+\frac{a_0^2}{r^2} \,.
    \label{metricfunc}
\end{equation}
In these functions, $r_+$ is the outer event horizon radius, $r_-$ is the radius of an inner Cauchy horizon, and $r_0$ is defined by $r_0=\sqrt{r_+r_-}$. These quantities are parametrised in the following way:
\begin{align}
&r_+=\frac{r_s}{(1+P)^2}\,, &r_-=&\frac{r_s P^2}{(1+P)^2}\,, &r_0=&\frac{r_s P}{(1+P)^2} \,,\\
&P=\frac{\sqrt{1+\epsilon^2}-1}{\sqrt{1+\epsilon^2}+1} \,, &\epsilon=&\gamma \delta \,.
\nonumber
\end{align}
Overall, the metric components depend on solely two parameters, say $a_0$ and $P$, which characterize the deviation of the polymerized BH from the Schwazrschild BH. The Schwarzschild black hole is recovered by taking $a_0$ and $P$ equal to zero.

The Barbera-Immirzi parameter $\gamma$ and the minimal area $a_0$ are constants from the fundamental LQG theory. To be precise, $a_0$ is the scaled version of LQG's area gap $\Delta$: $a_0=\Delta/8\pi$.
The adimensional polymeric parameter $\delta$ and the polymeric deformation function $P$ are effective parameters a priori resulting from the coarse-graining of the fundamental Planck scale dynamics. They are free parameters, to be determined by measurements. In fact, they are already constrained by astrophysical data \cite{Zhu:2020tcf,Yan:2022fkr,Liu:2023vfh}: $P < 6.17\times10^{-3}$ and $\delta < 0.67$, at 95\% confidence level.

The pole at $r=0$, and thereby the singularity, is removed by taking a non-vanishing minimal area $a_0> 0$. The parameter $P$ does not play any role in the singularity resolution, a quick computation allows to check that the  Kretschmann scalar is finite at $r=0$ and $P$ only affects the position of the outer and inner horizons.
Note nevertheless that the degeneracy $r_0=r_-=0$ when $P$ vanishes. This affect the pole structure of the metric components, which might have an impact on some calculations.

\subsection{Scalar perturbations}

\begin{figure}[!h]
     \centering
     \vspace*{3mm}
     \begin{subfigure}[b]{0.48\textwidth}
         \centering
         \includegraphics[width=0.9\textwidth]{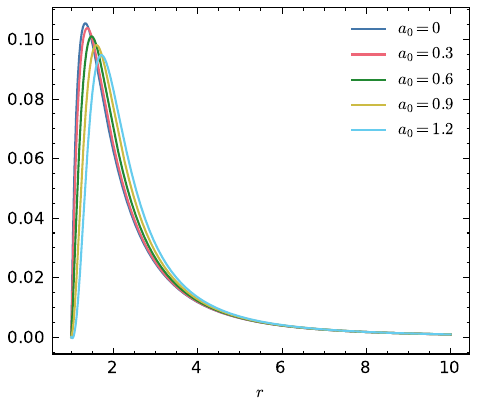}
         \caption{$P=0$}
         \label{pota_0}
     \end{subfigure}
     \hfill
     \begin{subfigure}[b]{0.48\textwidth}
         \centering
         \includegraphics[width=0.9\textwidth]{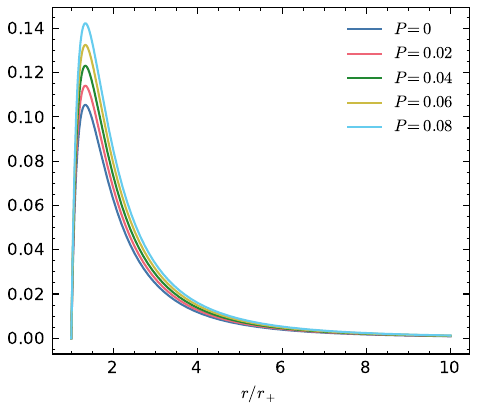}
         \caption{$a_0=0$}
         \label{potP}
     \end{subfigure}
     \caption{The effective potential for scalar perturbations as a function of the radial coordinate $r$ for $r_s = 1$, $l=0$, and different values of $a_0$ and $P$. The values have been chosen in order to best see the effects. It is a short ranged potential going to zero both at the horizon and at infinity. The dark blue curve with $a_0=P=0$ is the Regge-Wheeler potential for a Schwarzshild black hole.}
     \label{plotpot}
\end{figure}

We want to look at scalar perturbations in the background of the loop quantum BH described above in order to examine its scalar QNM spectrum. The dynamics of a massless scalar field $\Phi$ are given by the Klein-Gordon equation~\eqref{KG}. Similarly to the Schwarzschild case, we obtain a master equation for the function $\Psi(r)$:
\be 
\frac{\dd^2 \Psi}{\dd x^2} + [\omega^2-V(r)]\Psi = 0,
\label{schro}
\ee
with an effective potential including LQG corrections given by
\begin{align}
V(r)= \frac{1}{(a_0^2+r^4)^4} r^2(r-r_-)(r-r_+)\,\Big{[}
&10\,a_0^2\,r^4\,(r-r_-)(r-r_+)+a_0^4r(-2r+r_-+r_+)
\nn\\
&\,+ r^8[-2r_-r_+ +r(r_- +r_+)]+ \lambda\lp a_0^2+r^4\rp^2(r+r_0)^2\lambda\Big{]},
\label{pot}
\end{align}
%
%
where we called $\lambda=l(l+1)$ the angular momentum contribution. One recovers the Schwarzschild potential~\eqref{potSchwa} in the limit $a_0 = P = 0$.
We plot, on Figure \ref{plotpot}, the evolution of the effective potential \eqref{pot} with respect to the radial coordinate $r$ for different values of the parameters $a_0$ and $P$. The height of the effective potential decreases with $a_0$ while it increases with $P$. Moreover, the location of the maximum of the potential, corresponding to the light ring radius, increases with $a_0$ and slightly decreases with $P$. Finally, one notices that
the deformation $P$ seems to create larger deviation, with respect to the standard potential, than $a_0$, so that we expect $P$ to have a greater impact on the QNM values than $a_0$.

As in the Schwarzschild case, the effective potential vanishes on the horizon $r = r_+$  and at radial infinity $r \to +\infty$. We thus have asymptotic plane-waves in the tortoise coordinate $x$.
The tortoise coordinate is defined as earlier in~\eqref{eq:tortoise}, by the differential equation $\dd_rx=1/\sqrt{fg}$ in terms of the metric functions $f(r)$ and $g(r)$ now given by~\eqref{metricfunc}.
The behavior of  this tortoise coordinate $x$ is such that
\begin{subequations}
    \begin{align}
        &x = r + (r_+ + r_-) \ln(r) + \mathcal{O}(1) \,, &\mkern-120mu(r \to +\infty)\\
        &x = \sigma \ln(r - r_+) + \mathcal{O}(1)  \,, &\mkern-120mu(r \to r_+)
    \end{align}
    \label{eq:asymp-tortoise}
\end{subequations}
where
\begin{equation}
    \sigma = \frac{r_+^4 + a_0^2}{r_+^2 (r_+ - r_-)} \,.
\end{equation}
The computation of QNMs then requires imposing waves to be incoming at the BH outer horizon and outgoing at infinity,
%
%
\begin{subequations}
    \begin{align}
        &\Psi \sim e^{i\omega x} \sim e^{i\omega r}r^{i\omega(r_- +r_+)} \,, &\mkern-120mu(r \to +\infty)\\
        &\Psi \sim e^{-i\omega x} \sim (r-r_+)^{-i\omega\sigma}  \,. &\mkern-120mu(r \to r_+)
    \end{align}
    \label{asympto}
\end{subequations}
Let us now see how to translate this into recursion relations and continued fractions.

\subsection{Continued fraction method}

\begin{figure}[!h]
    \centering
    \includegraphics{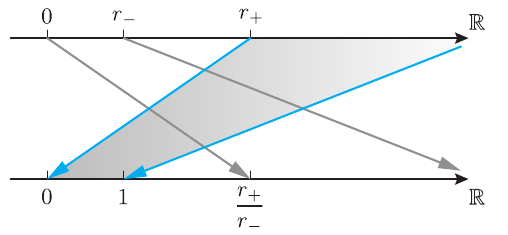}
    \caption{Representation of the mapping defined by equation \eqref{mapnew} with a highlighting of the values taken by the main singularities of equation \eqref{schro}, and of the domain of definition. This is inspired by the figure 3 in \cite{Moreira_2023}.}
    \label{map}
\end{figure}

Let us now turn to the computation of QNMs for the polymer BH using the continued fraction method reviewed in section~\ref{sec:contfrac}. The first step is to build an ansatz for the field $\Psi$ interpolating between the two boundary conditions~\eqref{asympto}. This can be done in the same way than for the Schwarzschild case, but adapting the mapping to the existence of both inner and outer horizons at $r=r_\pm$:
\be 
r \mapsto \frac{r-r_+}{r-r_-}.
\label{mapnew}
\ee
This sends the BH exterior onto the unit real interval, $(r_+,\infty)\,\to\,(0,1)$, as illustrated on figure \ref{map}. 

Then, taking into account the QNM boundary conditions \eqref{asympto}, we introduce an ansatz for the radial wave function solution of the master equation \eqref{schroSchwa}:
\begin{equation}
    \Psi(r)= e^{i\omega(r-r_+)}(r-r_-)^{i\omega(r_-+r_+)}\lp \frac{r-r_+}{r-r_-}\rp^{i\omega\sigma} \sum_{n=0}^{\infty}a_n\lp \frac{r-r_+}{r-r_-}\rp^n,
\label{ansatz}
\end{equation}
This ansatz correctly reduces to the Schwarzschild one~\eqref{eq:ansatz-schwarzschild} when  $a_0$ and $P$ are sent to 0.
After inserting this power expansion~\eqref{ansatz} in the Schr\"odinger equation~$\eqref{schro}$ for the scalar field, one obtains recursion equations relating the coefficients $a_n$ together. For instance, imposing that the leading order term in $(r-r_+)/(r-r_-)$ vanishes implies a relation between $a_0$ and $a_1$:
%
\begin{equation}
    c_{-1}(0) a_1 + c_0(0) a_0 = 0 \,.
\end{equation}
Setting $r_s=1$,  the coefficients $c_{-1}(0)$ and $c_0(0)$ explicitly read:
\begin{align}
c_{-1}(0)
=
(P+1)^2 \left(a_0^2 (P+1)^8+1\right)^2 \left[-(P-1) (P+1)^3-2 i \omega  \left(a_0^2 (P+1)^8+1\right)\right]\,,
\end{align}
\begin{align}
c_0(0)
=
&\big{[}-a_0^2 (P+1)^8-1\big{]} \Big{[}4 a_0^6 (P+1)^{24} \omega ^2-2 i a_0^4 P^2 (P+1)^{18} \omega +2 i a_0^4 (P+1)^{18} \omega -a_0^2 (\lambda +1) P^8 (P+1)^8
\\
&-2 a_0^2 (3 \lambda +2) P^7 (P+1)^8-2 a_0^2 (7 \lambda +2) P^6 (P+1)^8-2 a_0^2 (7 \lambda -2) P^5 (P+1)^8+2 a_0^2 P^4 (P+1)^8 (5+i \omega )
\nn\\
&+2 a_0^2 P^3 (P+1)^8 (7 \lambda +2 i \omega +2)+2 a_0^2 (7 \lambda -2) P^2 (P+1)^8+2 a_0^2 P (P+1)^8 (3 \lambda -2 i \omega -2)+a_0^2 \lambda  (P+1)^8
\nn\\
&-12 a_0^2 (P+1)^8 \omega ^2-2 i a_0^2 (P+1)^8 \omega -a_0^2 (P+1)^8+\lambda -\lambda  P^8+P^8-6 \lambda  P^7+4 P^7-14 \lambda  P^6+4 P^6-14 \lambda  P^5
\nn\\
&-4 P^5+4 i P^4 \omega -10 P^4+14 \lambda  P^3+8 i P^3 \omega -4 P^3+14 \lambda  P^2+4 P^2+6 \lambda  P-8 i P \omega
+4 P-8 \omega ^2-4 i \omega +1\Big{]}
\,.\nn
\end{align}
In general, the coefficients $a_n$ satisfy  a fifteen term recursion relation of the form
\begin{equation}
    \sum_{p = -1}^{13} c_p(n) a_{n-p} = 0 \,,
    \label{15termseq}
\end{equation}
where the recursion coefficients $c_p(n)$ (with the convention $c_p(n) = 0$ for $n < p$) depend on the BH parameters.
These coefficients are polynomials in the parameters $P$, $a_0$ and $\omega$, but their expression are long and unenlightening. They can be found in an attached Mathematica notebook \url{QNM_recursion_coefficients}.
%
One can show that the recursion relation~\eqref{15termseq} reduces when $a_0=P=0$ to the recursion of order 3  obtained for Schwarzschild BHs in \eqref{eq:rec-rel-schwa}.

It is possible to obtain a simpler recursion relation by scaling the radial wave function. Indeed, defining a rescaled wave function
\be
\Psi(r)=\frac{r}{\sqrt{h(r)}}\Xi(r)
\,,\quad
\Xi(r)=
e^{i\omega(r-r_+)}(r-r_-)^{i\omega(r_-+r_+)}\lp \frac{r-r_+}{r-r_-}\rp^{i\omega\sigma} \sum_{n=0}^{\infty}\ba_n\lp \frac{r-r_+}{r-r_-}\rp^n
\,.
\label{def-Xi}
\ee
$\Xi$ and $\Psi$ have the same asymptotic behaviors both at the horizon and at infinity, but the resulting recurrence relation now contains only seven terms~:
\begin{equation}
    \sum_{p = -1}^{5} \tilde{c}_p(n) \ba_{n-p} = 0 \,.
    \label{7termseq}
\end{equation}
The explicit formulas for the coefficients $\tilde{c}_p(n)$ can be found in an attached Mathematica notebook \url{QNM_recursion_coefficients}.
%
%
In order to extract the quasi-normal frequencies the recursion relations \eqref{7termseq} or~\eqref{15termseq}, one wishes to cast these recursion relations into simpler forms similar to the one obtained in the Schwarzschild case. This is possible by making use of the Gaussian reduction procedure, which we can apply to both 7-level an 15-level recursion relations. Let us define new coefficients $c_p^{(1)}(n)$ by
\be 
c_{-1}^{(1)}(n)= c_{-1}(n) \hspace{3cm}\text{for $n\geq0,$}
\ee
and for $1\leq p\leq p_{\text{max}}$:
\begin{equation}
    \left\{
        \begin{aligned}
        c^{(1)}_p(n)&= c_p(n) \hspace{5cm}\text{for $0\leq n\leq p_{\text{max}}-3$},\\
        c^{(1)}_p(n)&= c_p(n)-\frac{c^{(1)}_{p-1}(n-1)}{c^{(1)}_{p_{\text{max}}-1}(n-1)}c_{p_{\text{max}}}(n)\hspace{1cm}\text{for $n\geq p_{\text{max}}-2,$}
        \end{aligned}
    \right.
\end{equation}
where $p_{\text{max}}$ is the number of terms in the recursion relation --- 15 in the case of~\eqref{15termseq}. The effect of this procedure is to reduce the number of coefficients in the recursion relation by one. By iterating it $p_\mathrm{max} - 3$ times, one defines coefficients $c^{(2)}_p(n)$, $c^{(3)}_p(n)$..., and finally, the initial recursion relation can be cast into a three-term recurrence relation of the form
\begin{subequations}
    \begin{align}
        &\alpha(0)a_1+\beta(0)a_0=0,\label{2rec}\\
        &\alpha(n)a_{n+1}+\beta(n)a_n+\gamma(n)a_{n-1}=0 \ \ \ \ \text{for \ }n\geq 1.\label{3rec}
    \end{align}
    \label{3termeq}
\end{subequations}
Since this procedure is recursive, there is no general expression for $\alpha(n)$, $\beta(n)$ and $\gamma(n)$. Therefore, the expressions for $\alpha(n), \ \beta(n)$ and $\gamma(n)$ are very involved and it is necessary to evaluate them numerically. We follow the same procedure to reduce the level of the recursion relation driving the power expansion of the rescaled wave-function $\Xi(r)$.

For the ansatz \eqref{ansatz} to have a correct asymptotic behaviour, the series $\sum_{n}a_n$ must be convergent. As we  formulated the problem into a similar form to the Schwarzschild case, it is possible to use again Leaver's argument~\cite{Leaver:1985ax}: in order for the power series to converge, it is necessary for the coefficients $\alpha(n)$, $\beta(n)$ and $\gamma(n)$ to verify the \emph{continued fraction relation}~\eqref{eq:contfrac-schwa}.
The roots of this equation give the QNM frequencies.
Since the continued fraction is an infinite expansion, one needs to truncate it to a finite number of terms in order to compute QNMs numerically. We used between $N=200$ and $N=2000$ terms in our numerics depending on the desired precision. As a consistency check, we also checked that increasing the number of terms only improves the precision of the computed QNMs.

Moreover, one should note that it is possible to invert the first $n$ terms of the continued fractions and send them to the other side of the equation. This means that \eqref{eq:contfrac-schwa} can be re-written as:
\be 
\beta(n)-\frac{\alpha(n-1)\gamma(n)}{\beta(n-1)-}\frac{\alpha(n-2)\gamma(n-1)}{\beta(n-2)-}...-\frac{\alpha(0)\gamma(1)}{\beta(0)}=\frac{\alpha(n)\gamma(n+1)}{\beta(n+1)-}\frac{\alpha(n+1)\gamma(n+2)}{\beta(n+2)-}\frac{\alpha(n+2)\gamma(n+3)}{\beta(n+3)-}...,
\ee
As empirically observed in~\cite{Leaver:1985ax}, this reformulation of the equation has greater stability and optimal precision for the computation of the $n$th QNM. We therefore used this technique to improve the precision of our computations at high overtones.

Finally, as another consistency check, we made sure that the QNMs computed using the 15-terms recursion relation~\eqref{15termseq} are equal to the ones computed using the 7-terms recursion relation~\eqref{7termseq}. This comb allowed to eliminate spurious zeros from the continued fraction equation.
In the next section, we share and examine our results.

\section{Results}
\label{sec:results}

In the following section, we present our results, which highlight the main characteristics of the polymerised BH QNM spectrum.
We will set $r_s=1$ for all the numerics.

A  limitation of  the continued fraction method is its unreliability for purely imaginary frequencies \cite{Berti:2009kk,Moreira_2023,Daghigh_2023}. Indeed, as explained in more details in appendix \ref{app:imaginaryaxis},  zeros of the continued fraction equations are found all along the imaginary axe, but these are not stable with respect to the truncation $N$ and thus cannot be considered as QNMs. 
%
%
In the present state, our approach does appear to be reliable to detect purely imaginary QNMs. Such QNMs might in principle appear, as for example for de Sitter BHs in odd space-time dimensions \cite{Berti:2009kk}, but establishing their existence would require a different approach. In particular, a more careful analysis of the QNM boundary conditions seems to be required to characterize purely damped modes.


First we focus on the fundamental mode, \textit{i.e.} the least damped one, and analyze how its value is affected by the quantum parameters $a_0$ and $P$.
Then, we  extend this analysis to the first six QNM in order to understand how they get shifted when varying the two quantum parameters.
Finally, we go further down towards high imaginary part and look at the asymptotic behaviour of the QNMs at high overtone. In particular, while the spectra at $P\ne 0$ has a similar shape than Schwarzschild's, we surprisingly find a new crossing of the imaginary axis when the area gap is turned on, $a_0\ne0$, where the two QNM branches cross and the real part of their frequencies switches sign.
%

\subsection{Fundamental mode}

To begin with, let us study the behavior of the fundamental mode $\omega_0$, with the smallest imaginary part (in absolute value). This mode is particularly  relevant since it is the least damped one and thus the least suppressed physically.
We investigate to what extent this mode is affected by the change of the parameters $a_0$ and $P$.
To this purpose, we compute the complex frequency $\omega_0$ for different values of $a_0$ and $P$ and evaluate the relative difference with respect to the fundamental mode of the Schwarzschild black hole (\textit{i.e.} taking $a_0=P=0$). We realize a ten digits precision on the frequencies by appropriately adjusting the number of terms $N$ in the continued fraction. An analysis of the convergence of the continued fraction method with $N$ is given in appendix \ref{convergence}, confirming that the error is indeed lower than the target precision.
\begin{figure}[!h]
     \centering
     \begin{subfigure}[b]{0.48\textwidth}
         \centering       
         \includegraphics[width=0.9\textwidth]{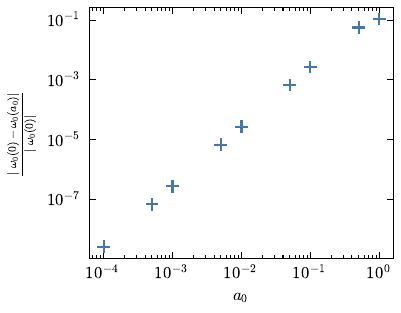}
     \end{subfigure}
     \hfill
     \begin{subfigure}[b]{0.48\textwidth}
         \centering
         \includegraphics[width=0.9\textwidth]{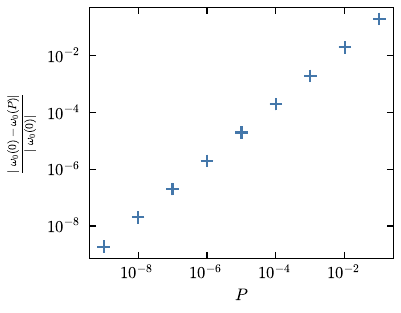}
     \end{subfigure}
     \caption{Evolution of the fundamental scalar mode $\omega_0$ at $l=0$ for the polymerized BH with respect to $a_0$ (on the left) and with respect to $P$ (on the right). The plot is showing the relative difference with respect to the Schwarzschild fundamental mode $\omega_0(a_0=0,P=0)=0.2209098782 - 0.2097914341 i$. }
    \label{plot_omega0_div}
\end{figure}

The results, for vanishing angular momentum $l=0$, are given in the plots of figure \ref{plot_omega0_div} and the detailed values of the frequencies are given by the tables \ref{omega0&P=0} and \ref{omega0&a0=0}.
%
%
The fundamental mode is more sensitive to $P$ than to $a_0$. More precisely, an absolute difference of about $10^{-4}$ between the Schwarzschild value and the modified value is obtained for either $a_0\simeq0.05$ or $P\simeq10^{-4}$. More generally, the effects of $a_0$ and $P$ on the frequency $\omega_0$ are cumulative, but here we chose to show the effects of each parameter separately for clarity. 


\vspace*{5mm}
\begin{minipage}[c]{0.5\textwidth}
\centering
    
    \begin{tabular}{|c||c|c|c|}
    \hline
    $a_0$  & $\omega_0$ \\
    \hline
    0 & $\textbf{0.2209098782} - \textbf{0.2097914341} i$ \\
    \hline
    0.0001 & $\textbf{0.22090987}73 - \textbf{0.20979143}39 i$ \\
    \hline
    0.0005 & $\textbf{0.2209098}562 - \textbf{0.2097914}272 i$ \\
    \hline
    0.001 & $\textbf{0.220909}7904 - \textbf{0.2097914}063 i$ \\
    \hline
    0.005 & $\textbf{0.22090}76846 - \textbf{0.20979}07377 i$ \\
    \hline
    0.01 & $\textbf{0.2209}011042 - \textbf{0.2097}886488 i$ \\
    \hline
    0.05 & $\textbf{0.220}6907093 - \textbf{0.2097}221211 i$ \\
    \hline
    0.1 & $\textbf{0.22}00354611 - \textbf{0.209}5181596 i$ \\
    \hline
    0.5 & $\textbf{0.2}008135535 - \textbf{0.20}57183839 i$ \\
    \hline
    1 & $\textbf{0.}1642192325 - \textbf{0.2}161031011 i$\\
    \hline
    \end{tabular}
    \captionof{table}{$P=0 \ (l=0)$}
    \label{omega0&P=0}

\end{minipage}
\begin{minipage}[c]{0.5\textwidth}
\centering

    \begin{tabular}{|c||c|c|c|}
    \hline
    $P$  & $\omega_0$ \\
    \hline
    0 & $\textbf{0.2209098782} - \textbf{0.2097914341} i$ \\
    \hline
    $10^{-9}$ & $\textbf{0.220909878}6 - \textbf{0.209791434}5 i$ \\
    \hline
    $10^{-8}$ & $\textbf{0.2209098}826 - \textbf{0.20979143}84 i$ \\
    \hline
    $10^{-7}$ & $\textbf{0.220909}9223 - \textbf{0.2097914}761 i$ \\
    \hline
    $10^{-6}$ & $\textbf{0.2209}103200 - \textbf{0.209791}8538 i$ \\
    \hline
    $10^{-5}$ & $\textbf{0.2209}142964 - \textbf{0.20979}56300 i$ \\
    \hline
    $10^{-4}$ & $\textbf{0.2209}540617 - \textbf{0.209}8333928 i$ \\
    \hline
    $10^{-3}$ & $\textbf{0.22}13518524  - \textbf{0.2}102110506 i$ \\
    \hline
    $0.01$ & $\textbf{0.22}53434087 - \textbf{0.2}139902957 i$ \\
    \hline
    $0.1$ & $\textbf{0.2}664963225 - \textbf{0.2}517228489 i$ \\
    \hline
    \end{tabular}
    \captionof{table}{$a_0=0 \ (l=0)$}
    \label{omega0&a0=0}

\end{minipage}

\subsection{Low overtone modes}

\begin{figure}[!h]
     \centering
     \begin{subfigure}[b]{0.48\textwidth}
         \centering
         \includegraphics[width=0.9\textwidth]{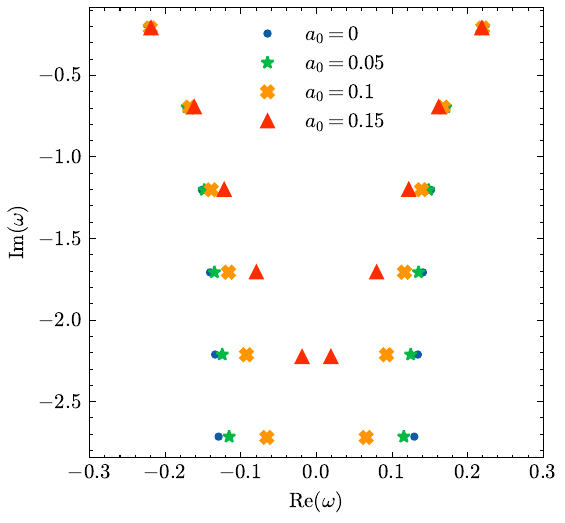}
         \caption{$l=0$}
     \end{subfigure}
     \hfill
     \begin{subfigure}[b]{0.48\textwidth}
         \centering
         \includegraphics[width=0.9\textwidth]{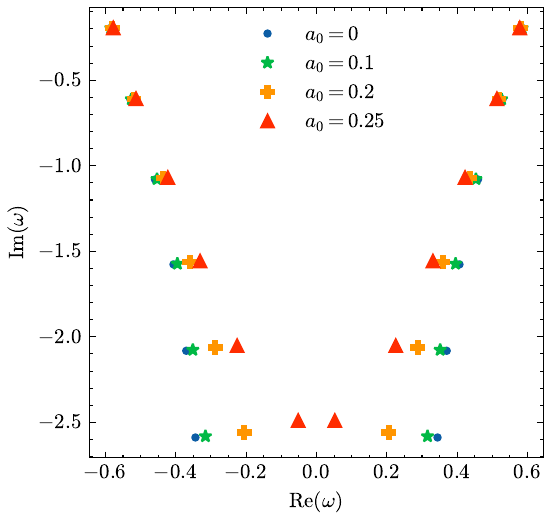}
         \caption{$l=1$}
     \end{subfigure}
     \caption{First six scalar QNM frequencies of the polymerized BH for $P=0$ and different values of $a_0$. The left spectrum is for $l=0$ and the right one for $l=1$.}
     \label{low_a0}
\end{figure}

\begin{figure}[!h]
     \centering
     \begin{subfigure}[b]{0.48\textwidth}
         \centering
         \includegraphics[width=0.9\textwidth]{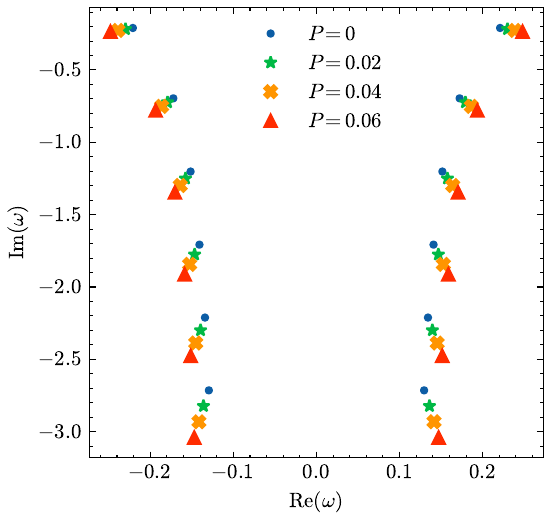}
         \caption{$l=0$}
         \label{}
     \end{subfigure}
     \hfill
     \begin{subfigure}[b]{0.48\textwidth}
         \centering
         \includegraphics[width=0.9\textwidth]{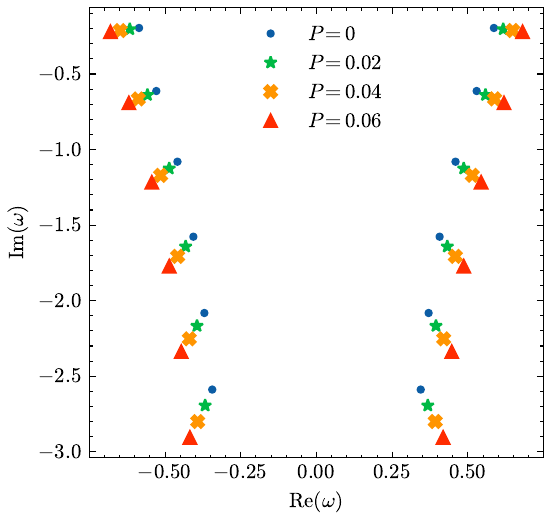}
         \caption{$l=1$}
         \label{}
     \end{subfigure}
     \caption{First six scalar QNM frequencies of the polymerized BH for $a_0=0$ and different values of $P$. The left spectrum is for $l=0$ and the right one for $l=1$.}
     \label{low_P}
\end{figure}

We now look at the low-damped part of the spectrum and analyse the effects of the parameters $a_0$ and $P$. More precisely, we compute the values of the six first modes for a range of values of $a_0$ and $P$, which clearly reflect the effects of the LQG deformation. We give the result for angular momenta $l=0$ and $l=1$. The results are displayed in figure \ref{low_a0} and \ref{low_P}.

To understand the physical impact of a non-vanishing area gap, $a_0\ne 0$, we set $P=0$ and vary the value of $a_0$.
On the one hand, the real part of every modes of the frequencies decreases (in absolute value) for both $l=0$ and $l=1$, as shown on figure \ref{low_a0}. On the other hand, the imaginary part slightly increases (in absolute value) for $l=0$, the modes thus getting more damped, while it slightly decreases (in absolute value) for $l=1$.
%
It is important to remark that the higher the overtone, the larger the deviation:
the larger $n$ is, the further the frequency $\omega_n$ deviates from the corresponding Schwarzschild value.
Moreover, although the spectrum  for small values of $a_0$ looks similar in shape to Schwarzschild's, it seems to give birth to intersecting branches, where the QNM frequencies cross the imaginary axis, for large enough values of $a_0$.
This holds for both $l=0$ and $l=1$, even if the $a_0$-threshold for this effect is larger for $l=1$ than for $l=0$.
%
%
This is rather surprising, since there is no crossing of branches for scalar perturbations (spin-0) in the Schwarzschild case. Actually, such crossing is a characteristic of gravitational perturbations (spin-2). This is thus a new feature of polymerized BHs.
This could have a deep theoretical origin, since a new crossing of the imaginary axis usually points to the existence of a QNM mode with vanishing real part, which appears to be symptomatic of a non-trivial symmetry \cite{Qi:1993ey}. This begs the question of whether polymerized BHs have less, equal or more symmetries than Schwarzschild BHs.

Let us now look at the deviations due to the polymeric parameter $P$, and set $a_0=0$ in the purpose of cleanly distinguishing the effects of the two deformation parameters.
Unlike for $a_0$, the first mode already undergoes an important shift in real part. All the modes increases (in absolute value) in both real and imaginary parts, the larger the overtone number $n$ is, the larger the deviation of $\omega_n$ becomes. Nevertheless, the overall shape of the spectrum for polymerized BHs at $a_0=0$ remains similar to the Schwarzschild spectrum, with no surprising new feature.
This clearly illustrates the different effect and role of the area gap $a_0$ and polymeric deformation parameter $P$.

\subsection{High Imaginary part}


We push our analysis further and dive deeper in the complex plane towards frequencies with higher imaginary parts, in order to 
investigate the deviation of polymerized QNMs in the highly-damped part of the spectrum. 

\begin{figure}
    \centering
    \includegraphics[width=0.5\textwidth]{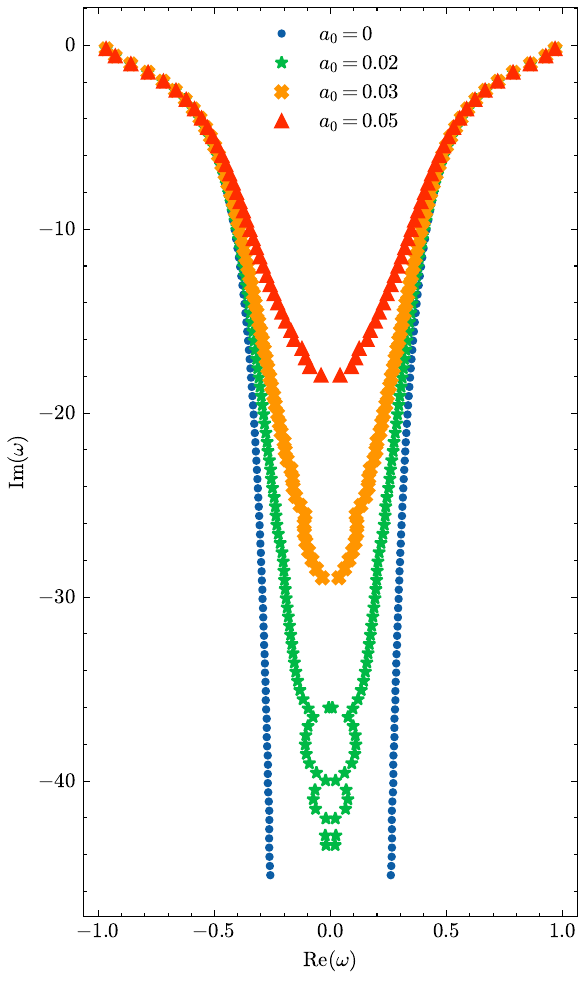}
    \caption{QNM spectra for the polymerized BH computed for different values of $a_0$ and $l=2$, at $N=1500$ for the truncation of the continued fraction.}
    \label{big_a0}
\end{figure}

First, let us look at the evolution of the spectrum with the area gap $a_0$. Setting $P=0$, we see that the QNM spectrum becomes qualitatively different from the Schwarzschild case as we increase the value of $a_0$.
Indeed, as we already noticed in our analysis of low-damped modes, we see that, for sufficiently large values of $a_0$, we see on figure \ref{big_a0} that the two QNM branches meet and cross the imaginary axis. For smaller values of $a_0$, the numerics seem to indicate that the crossing probably still occurs but further along the imaginary axis. Checking this would require higher precision at high imaginary part.
Let us underline three artefacts of our method:
\begin{itemize}
\item
Before the crossing, one can spot small oscillating patterns at high-imaginary part, for instance for $a_0=0.02$ and $a_0=0.03$ on figure \ref{big_a0}. Wondering if  these oscillations have a physical origin as in \cite{Moreira_2023}, we found that these are numerical errors due to a lack of precision, which can be suppressed by going to higher $N$'s in the computation of the continued fraction, as we analyse in detail in appendix \ref{osc_appendix}.

\item After the crossing, our code failed to identify clear QNMs after this crossing. Indeed, the frequencies released by our code were by far too numerous and erratic and our data analysis didn't make possible to extract clear stable QNM branches after the crossing.

\item We detect zeros of the continued fraction equation along the imaginary axis. As we have already underlined, these zeros are not stable under refinement of the continued fraction, as one increases the truncation level $N$, and we do not identify them as proper QNMs. These feature seems to be an intrinsic limitation of Leaver's method. As we explain in appendix \ref{app:imaginaryaxis}, these purely imaginary zeros are not a property of  polymerized BHs, but are generic and, in particular, already occur for the Schwarzschild metric.

\end{itemize}
In contrast to these three subtleties, the new crossing clearly emerges as a robust numerical prediction.
This crossing occurs sooner (i.e. for smaller imaginary part in absolute value) as the area gap $a_0$ increases. 
%
%
Such a crossing usually comes together with a QNM frequency with very small or vanishing real part, which is symptomatic of the existence of a hidden symmetry \cite{Qi:1993ey} and would signal new physics for polymerized BHs.

Let us nevertheless remark that, even if confirmed, such effects are not expected to be physically relevant for large astrophysical BHs, since the area gap $a_0$ is supposed to be of Planck scale. It could however impact the physics of Planck scale BHs, e.g. primordial BHs, and more particularly black hole - white hole (BH-WH) oscillations and their stability \cite{Rovelli:2018okm}.

\begin{figure}
    \centering
    \includegraphics[width=0.5\textwidth]{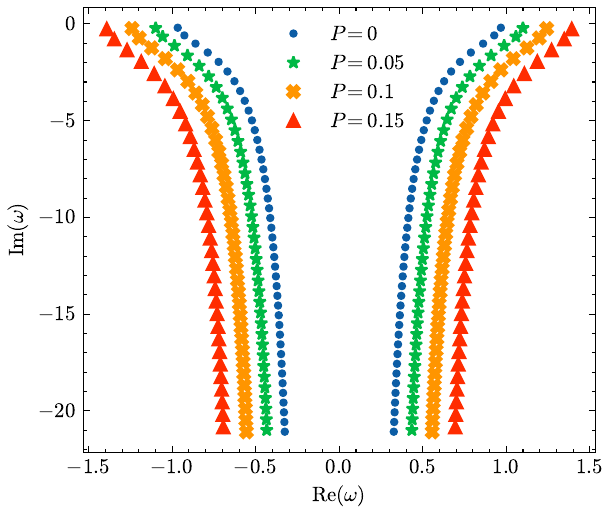}
    \caption{QNM spectra for the polymerized BH computed for different values of $P$ and $l=2$.}
    \label{big_P}
\end{figure}

\medskip

We can also set $a_0=0$ and study  the effect of a non-vanishing polymeric deformation parameter $P$.
We computed the QNM frequencies up to an imaginary part of around $-20i$, for a range of values of $P$. The results are plotted in figure \ref{big_P}.
The real part of the modes increases in absolute value with the polyemric deformation $P$.
This effect was already observed in the previous section for the first few QNMs and see here that it is a systematic deviation for all QNMs.
There is also a substantial shift in the imaginary part. It is less visible to the naked eye on figure \ref{big_P}, but, in fact, 43 modes are displayed for the Schwarzshild spectrum at $P=0$ while only 33 ones appears in the same range in the $P=0.15$ spectrum.
From here, it is interesting to push our analysis of the asymptotic of the QNMs.
Indeed, it is known for the Schwarzschild BH, in the highly-damped regime of perturbations, that the real part of the QNM frequencies converges towards the constant value $(\ln 3)/4\pi$, as shown analytically in \cite{Motl:2002hd,Motl:2003cd}, while the gap in imaginary part between two succeeding QNMs converges to a value of around $\frac{1}{2}$. Let us look at  the changes induced by a non-zero value of $P$.

\begin{figure}[!h]
     \centering
     \begin{subfigure}[b]{0.45\textwidth}
         \centering
         \includegraphics[width=0.85\textwidth]{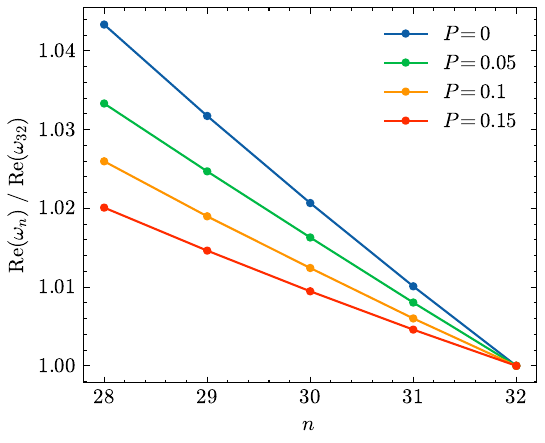}
         \caption{Evolution of the real part of the QNM at large $n$, for $n=28,..,32$, scaled with the real part of $\omega_{32}$.}
         \label{asympRe}
     \end{subfigure}
     \hspace*{5mm}
     \begin{subfigure}[b]{0.45\textwidth}
         \centering
         \includegraphics[width=0.85\textwidth]{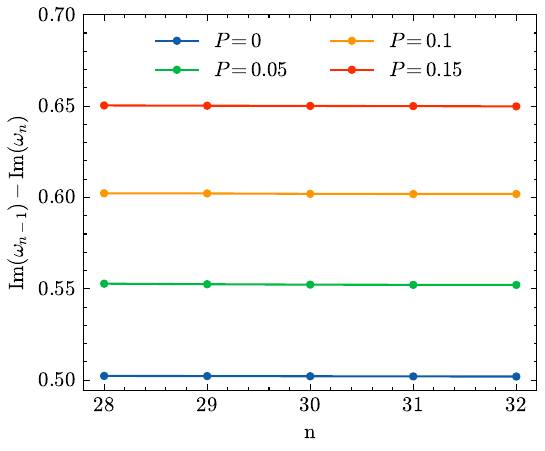}
         \caption{Evolution of the gap in imaginary part of successive QNM frequencies, at large $n$, for $n=28,..,32$.}
         \label{asympIm}
     \end{subfigure}
     \label{bigpot}
     \caption{Analysis of the asymptotic QNM frequencies and their dependence on $P$,  for  the real part (\textit{on the left}) and  the imaginary part (\textit{on the right}).}
\end{figure}

First, we look at  the evolution of the real part.
We plot the ratio between the real part of successive highly-damped QNMs for various values of the polymeric deformation parameter $P$ in figure \ref{asympRe}. 
One realizes that the ratios are smaller as  $P$ increases, meaning that the convergence is faster. Nevertheless, our code does not achieve sufficient precision for larger $n$'s in order to make a prediction for the limit real part and how it precisely depends on $P$, although it looks like it a linear growth at first sight.


Second, we look at the evolution of the imaginary part of the QNM and, more specifically, at the evolution of the gap between the imaginary parts of successive QNMs.
Numerics are plotted on figure \ref{asympIm}. One checks that the gap for Schwarzschild QNM is equal to 0.5 at leading order, as expected. For non-zero $P$, the gap also converges to a finite constant. At leading order, this gap seems to evolve linearly with $P$,
\be
\text{Im}(\omega_{n-1})-\text{Im}(\omega_n) \approx \f12+P.
\ee
Since this gap is supposed to probe the deep quantum regime of BHs \cite{Dreyer:2002vy,Konoplya:2011qq}, it would be of great interest to go further in the spectrum and check numerically this conjectured behavior. Unfortunately, we have reached the precision limit of our Mathematica code.
It would also be enlightening to compute the QNM asymptotic gap for polymerized BHs analytically, using for instance the monodromy method developed in \cite{Motl:2003cd}. This would provide deep insight into the quantum regime behavior of polymerized BHs and perhaps show if it is indeed consistent with the LQG's Planck scale dynamics.

\subsection{Correspondence with circular null geodesics in the eikonal limit}

In this section, we focus on the eikonal limit where $l\rightarrow \infty$ and we show that the QNM frequencies $\omega$ can be expressed in terms of the parameters of the circular null geodesics. Indeed, let us write $\Omega_c$ the coordinate angular velocity for a circular null geodesic and $\lambda$ the Lyapunov exponent, which gives the instability timescale of the geodesic. We find that the QNM frequencies can be approximated in the eikonal limit by:
\be
\omega \approx \lp l+\frac{1}{2}\rp \Omega_c -\lp n+\frac{1}{2}\rp \lambda.
\ee
As showed in full detail in appendix \ref{eikonal_appendix}, this formula holds for any spherically symmetric metric of the form \eqref{metric}, and the parameters $\Omega_c$ and $\lambda$ can be written in terms of the metric functions,
\be
\Omega_c = \sqrt{\frac{f_c}{h_c}}
\,,\qquad
\lambda=\sqrt{\frac{g_c}{2h_c}\lc f_ch_c''-h_cf_c''\rc}
\,,
\ee
where $f_c$, $g_c$ and $h_c$ are the values of the metric components $f(r)$, $g(r)$ and $h(r)$ at the radius of a circular null geodesic $r=r_c$. As also shown in appendix \ref{eikonal_appendix}, this QNM spectrum formula is equivalent at leading order with the WKB approximation.

\begin{figure}[!h]
    \centering
    \begin{subfigure}[b]{0.45\textwidth}
    \centering  \includegraphics[width=0.9\textwidth]{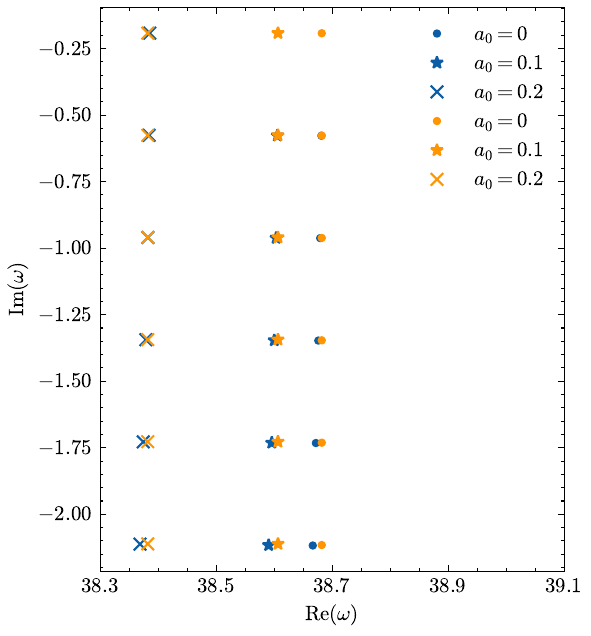}
    \caption{$P=0$}
    \label{}
    \end{subfigure}
    \hfill
    \begin{subfigure}[b]{0.45\textwidth}
    \centering
    \includegraphics[width=0.9\textwidth]{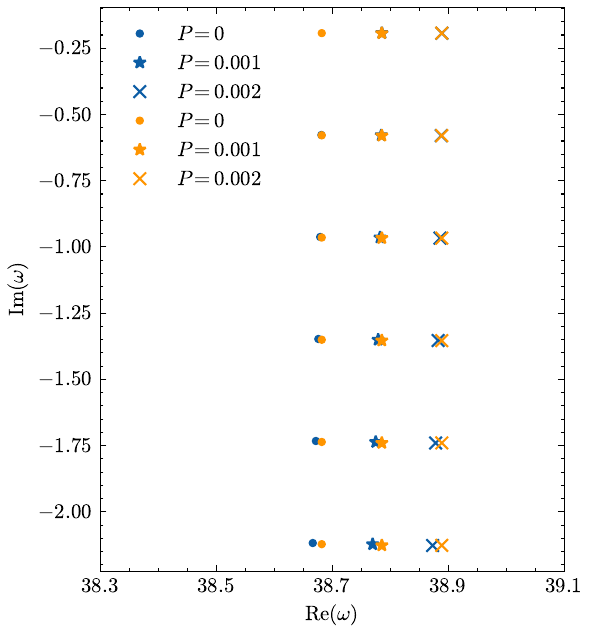}
    \caption{$a_0=0$}
    \label{}
    \end{subfigure}
    \caption{Evolution of the six first QNM with $a_0$ and $P$ in the eikonal limit ($l=100$).The blue points correspond to the QNM computed using Leaver's method, and the orange points to the leading order analytical prediction \eqref{eikonalQNM0}. }
    \label{eikonal_a0}
\end{figure}

Now we can check this result numerically by comparing the QNM frequencies obtained via Leaver's method and the analytical prediction given above in equation \eqref{eikonalQNM0}. The plots for the six first modes in figure \ref{eikonal_a0} correspond at an angular momentum $l=100$ and ranges of quantum parameters, $a_0\in[0,0.2]$ and $P\in[0,0.002]$. Explicit numerics are given in appendix \ref{app:numerics-eikonal}.

The agreement between the two sets of values not only validates the circular null geodesic approximation for the eikonal limit, but also confirms the robustness and accuracy of our implementation of Leaver's method.

%

\section*{Discussion and conclusion}

The primary purpose of the computation of Quasi-Normal Modes (QNMs) is to study the response of black holes (BHs) to  perturbations and describe their return to equilibrium dynamics.
Originally studied to prove the stability of BH solutions in General Relativity, QNMs regained a lot of interest over the last years with on the one hand the dawn of gravitational wave astronomy and BH spectroscopy.
Indeed, the quest for a quantum theory of gravity and enhanced cosmological models has led to the proliferation of effective models of modified BHs. As BH footprints, QNM are  expected to be a  good tool to characterise and discriminate effective models, in the hope that we would one day be able to observe a large part of the QNM spectrum and notice deviations from General Relativity. 

In the present paper, we apply this logic to BH metrics derived from Loop Quantum Gravity (LQG). There  indeed exists several proposals for modified BHs taking into account LQG corrections and insights (see appendix \ref{app:catalogue} for a catalogue of the main models). They all resolve the BH singularity, as one naturally expects from a quantum gravity theory, but nevertheless propose different geometries, especially for the BH interior. Thus the natural question of discriminating between them and, in particular, can QNMs reflect the geometry of the BH interior and the type of singularity resolution? 

Here we investigate one of the first effective BH models derived from LQG. Constructed by Modesto \cite{Modesto:2008im}, this polymerized BH accounts for LQG effects through two parameters: the area gap $a_0$ and the polymeric function $P$.
We perform a numerical investigation of the QNMs using Leaver's method and analyze the effects of each parameter on the spectrum. 
Among all the techniques which have been developed for the QNM computation, Leaver's  method, based on a reformulation of the wave equation in terms of continued fractions, distinguished itself by its accuracy and its efficiency on a large part of the QNM spectrum \cite{franchini2023testing}.
We indeed checked the accuracy of our results by checking its efficiency in the eikonal limit (large angular momentum) where an analytical asymptotic formula is known, by comparing them against other methods (WKB approximation and spectral decomposition, as shown in appendix \ref{app:compare}), and by comparing our numerics with previous work \cite{Liu:2020ola,Momennia:2022tug,Moreira_2023}, as detailed in appendix \ref{app:others}.
Finally, we ensured the validity of our approximation, especially the truncation of the continued fractions, by checking the stability of the numerics under refinements of the truncation, as explained in appendix \ref{convergence}.

\begin{figure}[!h]
    \centering
    \includegraphics[width=7cm]{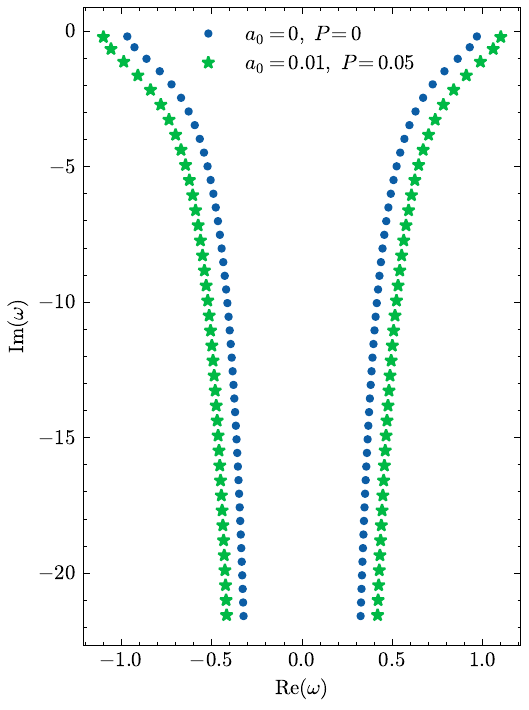}
    \caption{An example of spectrum for Modesto's polymerized BH, given by the metric components \eqref{metricfunc}, where both parameters are taken non-zero: $a_0=0.01, P=0.05$. The blue dots represent Schwarzshild spectrum as a reference. }
    \label{finalplot}
\end{figure}

We presented results for Modesto's polymerized BHs on the QNM frequency deviation from the standard Schwarzschild case, focusing on each parameter, $a_0$ and $P$ separately, in order to clearly distinguish their relevance and impact.
Therefore we presented plots and numerics for polymerized BHs where we alternately turned one of these two parameters off. Our code nevertheless perfectly accommodates the general case and can compute the QNMs when both parameters are non-zero, as illustrated on figure \ref{finalplot}.

First, focussing on the fundamental mode, which is the least damped and thus the most physically-important, we found that both quantum parameters shift its frequency, and that its value is especially sensitive to the polymeric deformation parameter $P$. To compare with 
the astrophysical constraints discussed in
\cite{Yan:2022fkr,Liu:2023vfh}\footnotemark,
and the upper bound found with the larger confidence level $0<P<6.17\times10^{-3}$, 
we get a notable difference of the fundamental frequency from the Schwarzschild value from the third decimal for $P\sim 10^{-3}$ (as shown on figure \ref{plot_omega0_div} and table \ref{omega0&a0=0}).
%
\footnotetext{
Earlier work \cite{Zhu:2020tcf}, by one of the authors of \cite{Yan:2022fkr,Liu:2023vfh}, actually obtained more stringent constaints, based on several astrophysical tests, among which gravitational time delay (by the Cassini experiment), perihelion advance and geodesic precession, yielding $0<P<5.5\times10^{-6}$. It is not entirely clear how this fits with the results presented later in \cite{Yan:2022fkr,Liu:2023vfh}, but this would imply a much smaller deviation of the fundamental QNM frequency.
}

The deviation for higher overtones is even more striking.
First, turning on $P$ increases both real and imaginary parts of the frequencies. The higher the overtone is, the larger the deviation is. This results in a faster convergence towards the asymptotics. Although we were not able to make a precise enough result for the asymptotical real part of the frequencies, our method was precise enough to evaluate the gap in imaginary part between successive QNMs. We confirmed that the spectrum is asymptotically equidistant, as for Schwarzschild BHs, and we conjecture an asymptotic gap of $\frac12+P$  linearly depending on the polymeric deformation.

Second, the effect of the area gap seems to be  more drastic. We see that, for non-vanishing values $a_0$, the two branches of QNMs cross. This property is absent for scalar perturbations of the pure Schwarzschild BH. In practice, this means that there exist QNMs whose real part is very close to vanishing. This feature is symptomatic of the existence of a hidden symmetry \cite{Qi:1993ey} and is related to the so-called \enquote{spectral instability} of BHs~\cite{Jaramillo:2020tuu,Jaramillo:2021tmt,Destounis:2021lum}.
Moreover, our code failed to clearly identify QNM branches beyond that crossing. This hints to the breakdown of Leaver's method, or of our implementation of it, beyond that critical QNM. This could perhaps due to QNM frequencies with very small real part, or the existence of multiple QNM branches, or a change in QNM boundary conditions, or some other reason all together.
In all cases, this signals new QNM physics for the high overtones of polymerized BHs.

These qualitative and quantitative modifications of the QNM spectrum arising from LQG's effetive polymerisation seems promising for the future of BH spectroscopy. We note nevertheless that we have been studying a scalar field propagating on the modified BH metric, and that we haven't yet investigated true BH perturbations based on a modified GR dynamics accounting for LQG effects on the Einstein equations. This underlines the necessity of the research effort in extracting effective modified gravity scenarii from LQG beyond spherical symmetry. This is even more crucial in light of the recently underlined relevance of non-linear dynamics for QNMs \cite{Yi:2024elj}.

We would like to conclude this paper with an apparent paradox about Polymerized Black Holes. It was underlined in previous litterature that the high overtones of the QNMs (for large imaginary parts) probe and reveal the quantumness of space-time and are apparently consistent with LQG's prediction of a discrete area spectrum \cite{Dreyer:2002vy}. However, we see here that the QNM deviation for polymerized BHs grows with the overtone number, which thus seems to imply a notable deviation from LQG's area spectrum. So are polymerized BHs, a priori derived from LQG, actually consistent with LQG? Or does this signal a deeper problem in modelling BHs in LQG? Or does this simply mean that the relation between high overtone QNMs and quantum gravity is more subtle than previously thought?

\acknowledgments{
We are grateful to Jibril Ben Achour and Che-Yu Chen for stimulating discussions on quasi-normal modes and black holes in modified gravity, and their encouragements in completing this work.

C.M. would like to also thank Carlos Herdeiro for interesting discussions on quasi-normal modes during the International Conference on Gravitation and Cosmology 2024, in Lahore, Pakistan.

N.O. is supported by the Grant-in-Aid for Scientific Research (JSPS KAKENHI) project (Grant Numbers JP23K13111) and by the Hakubi project at Kyoto University.
}

\appendix

\section{Catalogue of effective BH metrics from LQG}
\label{catalogue}
\label{app:catalogue}

We present a catalogue of the main BH effective metrics derived from LQG, constructed over the past few years, by considering different polymerisation schemes and different modifications of the Hamiltonian constraint.
All these modified Schwarzschild metrics fit in the general ansatz
\be
\dd s^2= -f(r)\dd t^2+\frac{\dd r^2}{g(r)}+h(r)\dd \Omega^2
\,,
\ee
and are defined through different prescriptions for the metric components $f(r)$, $g(r)$ and $h(r)$. Here are the main models and their features:

\begin{itemize}
    \item \textbf{Modesto metric} \cite{Modesto:2008im}: The original polymerized BH metric, which we study in the present work, is given by:
    \be
    f(r)=\frac{(r-r_+)(r-r_-)}{r^4+a_0^2}(r+r_0)^2,
    \ee
    \be
    g(r)=\frac{(r-r_+)(r-r_-)}{r^4+a_0^2}\frac{r^4}{(r+r_0)^2},
    \ee
    \be h(r)=r^2+\frac{a_0^2}{r^2}.\ee

    Several QNM computations have already been performed:
    \begin{itemize}
        \item first gravitational QNMs with $a_0$ set to 1, computed using the 6$^{th}$ order WKB method in \cite{moulin2019overview},
        \item first scalar and electromagnetic QNMs computed with the 6$^{th}$ order WKB technique, along with the construction of the corresponding rotating BH metric, in \cite{PhysRevD.101.084001},
        \item first scalar and electromagnetic QNMs computed thanks to the 6$^{th}$ order WKB technique and the asymptotic iteration method  in \cite{PhysRevD.106.024052},
        \item first axial gravitational QNMs, computed using WKB technique along with the asymptotic iteration method and restraining to $a_0=0$, in \cite{Yang_2023}.
    \end{itemize}

    \item \textbf{ABBV metric} \cite{Alonso_Bardaji_2022}: a new Loop Quantum BH metric constructed in 2022. The metric is characterised by a single quantum parameter, $r_0$, representing the radius of the minimal surface. 

    \be
    f(r)=\frac{r-r_s}{r},\ee
    \be g(r)=\frac{r-r_0}{r}f(r),\ee
    \be h(r)=r^2.\ee

    Some QNM computations have already been performed:
    \begin{itemize}
        \item about 30 scalar QNM overtones were computed using the third WKB approximation, the continued fraction method and the Prony method in \cite{Moreira_2023}. Interesting oscilating patterns were found.
        \item up to 9 scalar, electromagnetic and dirac QNM overtones were computed using the continued fraction method and the sixth order WKB approximation, in \cite{bolokhov2023longlived},
        \item up to 5 scalar and electromagnetic QNM overtones computed thanks to the pseudo-spectral method, in \cite{fu2023peculiar}.
    \end{itemize}
    
    \item \textbf{PK metric} \cite{Peltola_2009a}: an effective polymerised BH metric found in 2009. The metric is characterised by a single quantum parameter, $\alpha$, representing the radius of the minimal surface. 

    \be
    f(r)=\frac{r-r_s}{\sqrt{r^2+\alpha^2}},\ee
    \be g(r)=f(r),\ee
    \be h(r)=r^2+\alpha^2.\ee

    The scalar and electromagnetic fundamental QNMs were computed using the 6th order WKB technique in \cite{jha2023shadow}. In addition to that gravitational lensing and shadow were investigated in \cite{Walia_2023}, along with constraints for the quantum parameter.

    \item \textbf{BMM metric} \cite{Bodendorfer_2021}: a recent polymerised BH metric, depending on a single quantum parameter $A_{\lambda}$.

    \be
    f(r)=\frac{\sqrt{2A_\lambda r_s^2+r^2}(\sqrt{2 A_\lambda r_s^2+r^2}-r_s)}{\frac{1}{2} A_\lambda  r_s^2 +r^2},\ee
    \be g(r)=f(r),\ee
    \be h(r)=r^2+\frac{1}{2}A_\lambda r_s^2.\ee
    No QNM computation was ever done for this BH metric, but here is some work which was done about this BH:
    \begin{itemize}
        \item a corresponding rotating BH was constructed in \cite{Brahma_2021},
        \item  the frequency shifts of photons emitted by massive particles revolving around this non-rotating BH was studied in \cite{Fu_2023},
        \item gravitational lensing was investigated by \cite{PhysRevD.105.064020}.
    \end{itemize}
\end{itemize}

\section{Breakdown of the continued fraction method on the imaginary axis}
\label{app:imaginaryaxis}

\begin{figure}[!h]
     \centering
     \begin{subfigure}[h]{0.45\textwidth}
         \centering
         \includegraphics[height=65mm]{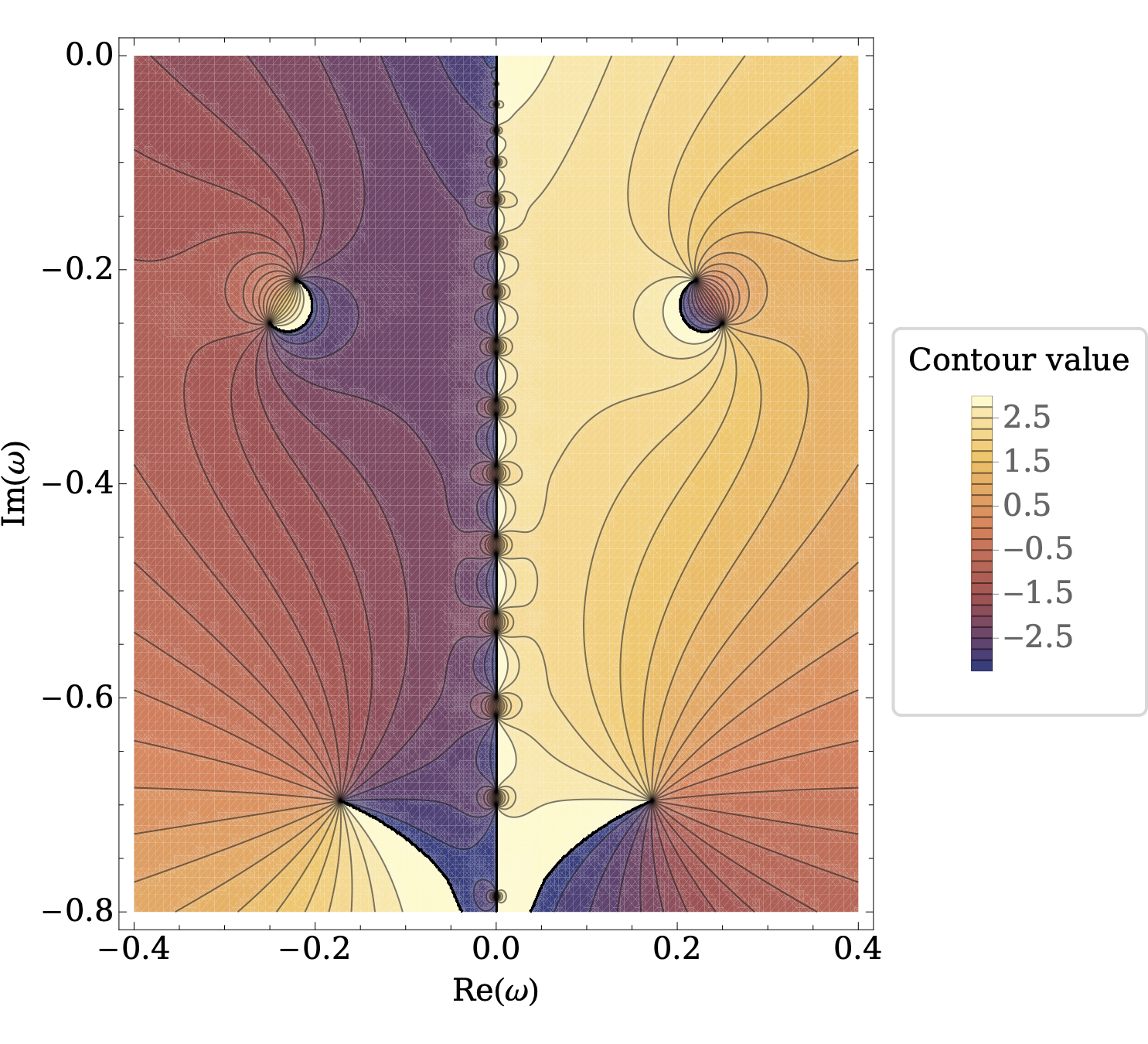}
         \caption{
         Contour plot of the argument of the continued fraction on the complex plane. The continued fraction was truncated at   $N=500$ and inverted once.
Vortices where the contour value increases counterclockwise give zeros, while vortices where it increases clockwise indicate poles.
         }
         \label{comp_Sch}
     \end{subfigure}
     \hspace*{8mm}
     \begin{subfigure}[h]{0.45\textwidth}
         \centering
         \includegraphics[height=65mm]{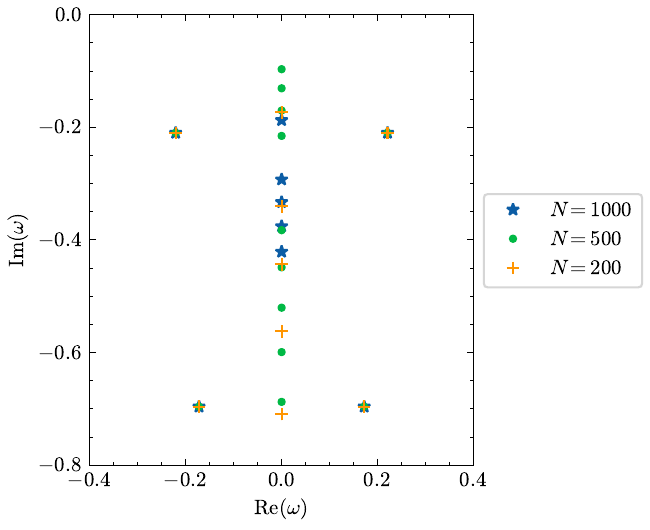}
         \caption{Zeros of the continued fraction for a Schwarzshild BH ($a_0=P=0$) found by our code for various values of the truncation $N$. Values of true QNMs are stable as $N$ increases while  zeros on the imaginary axe are not.}
         \label{Re0_Sch}
     \end{subfigure}
     \label{Sch_plot}
     \caption{Identifying Schwarzschild QNMs from continued fraction zeros.}
\end{figure}

In this section, we highlight the  behaviour of the continued fraction on the imaginary axe.
As one can see on figure \ref{Re0_Sch}, a whole sequence of zeros of the continued fraction for the polymerized black hole are found along the imaginary axe. These zeros are not stable under refinement as the truncation level $N$ increases and we do not identify them as proper QNMs.
\begin{figure}[!h]
     \centering
         \includegraphics[height=100mm]{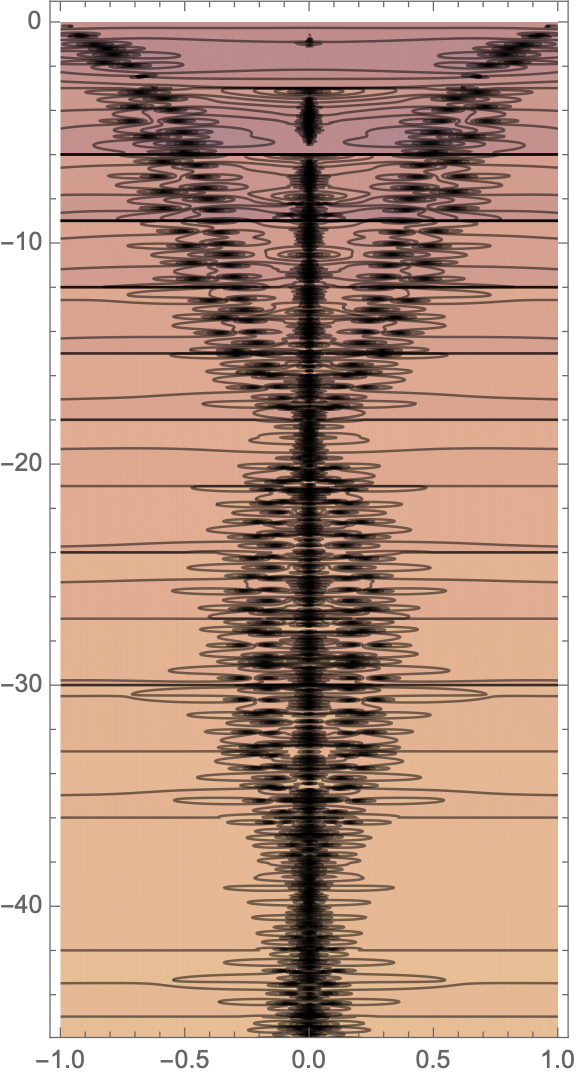}
         \includegraphics[height=50mm]{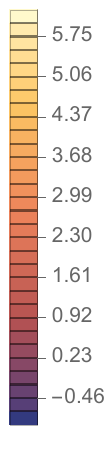}
         
     \caption{
Contour plot of the continued fraction with $N=2400$, for the polymerized BH with quantum parameters $a_0=0.05$, $P=0$ and angular momentum $l=2$.
The darker zones indicate zeros. Analysis is refined in those zones to extract precise values of these zeros.
Once we obtain the precise location of a zero, we check its existence for both the 7-term and 15-term recursions, its stability under changes in the truncation level $N$ and the convergence rate of the continued fractions. This allows to identify clear-cut QNMs and discard, for instance, the zeros along the imaginary axis.
%
Our QNM results are the remaining stable branches,
e.g. in figure \ref{big_a0}.}
     \label{fig:after}
\end{figure}

Actually, this is not a feature of the polymerized metric, but a general behavior of the continued fraction method. Indeed, as shown on  figure \ref{Re0_Sch}, the same behavior occurs for the standard Schwarzschild BH. Zeros are found on the imaginary axe. However, these  purely damped modes are not stable as we change the truncation $N$ of the continued fraction, unlike the four modes with non-zero real part shown on the same figure. These four modes are identified as real QNMs of the Schwarzschild BH in contrast to the other ones. This "breakdown" of the continued fraction method occurs for general BH metrics and persists in spite of very high precision. 
%

This stresses that one should be careful with the interpretation of purely imaginary zeros as QNMs.
Despite being pointed out in the review \cite{Berti:2009kk}, we are not aware of a clear-cut mathematical explanation for this behaviour of the continued fraction method.
A serious possibility could be the non-adequacy of the QNM boundary conditions, used here, for complex frequencies with vanishing real part. 

\section{Small oscillations: physically relevant or lack of precision?}
\label{osc_appendix}

\begin{figure}[!h]
     \centering
     \begin{subfigure}[b]{0.3\textwidth}
         \centering
         \includegraphics[width=\textwidth]{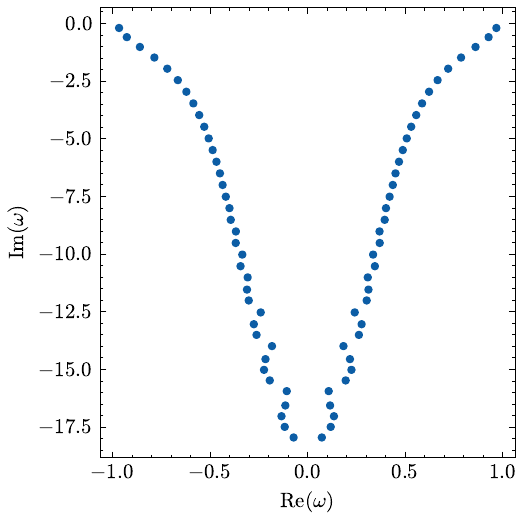}
         \caption{$N=200$}
         \label{}
     \end{subfigure}
     \begin{subfigure}[b]{0.3\textwidth}
         \centering
         \includegraphics[width=\textwidth]{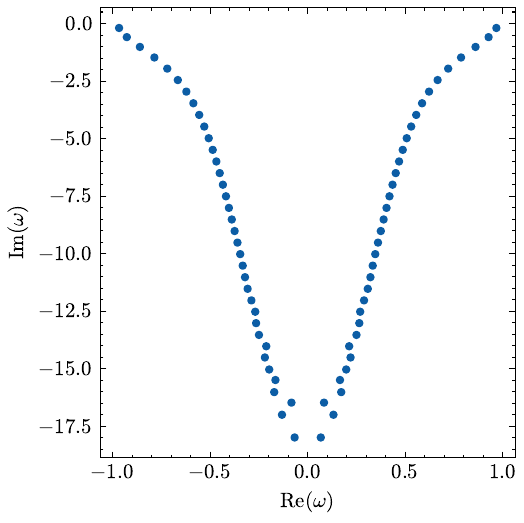}
         \caption{$N=400$}
         \label{}
     \end{subfigure}
     \begin{subfigure}[b]{0.3\textwidth}
         \centering
         \includegraphics[width=\textwidth]{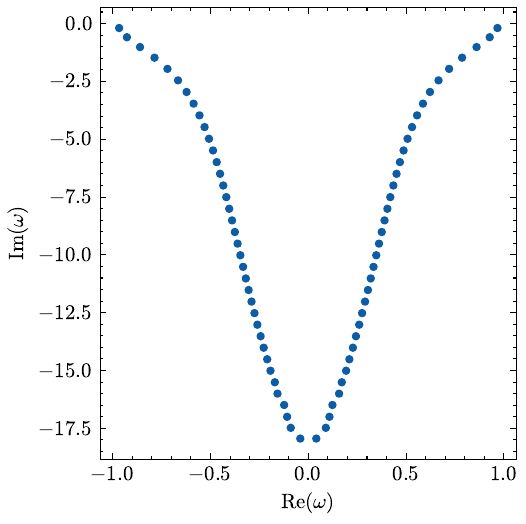}
         \caption{$N=800$}
         \label{}
     \end{subfigure}
     \hfill
     \begin{subfigure}[b]{0.3\textwidth}
         \centering
         \includegraphics[width=\textwidth]{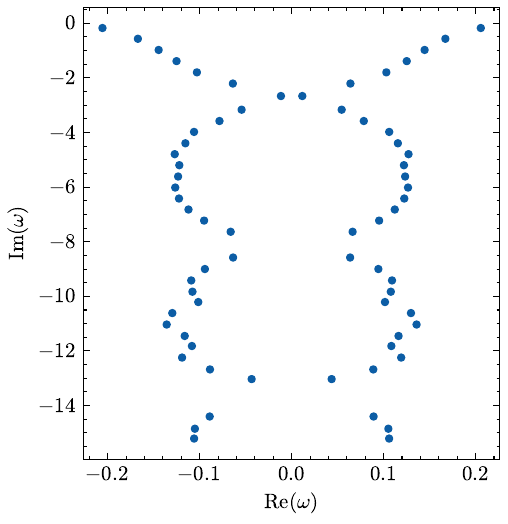}
         \caption{$N=800$}
         \label{}
     \end{subfigure}
     \begin{subfigure}[b]{0.3\textwidth}
         \centering
         \includegraphics[width=\textwidth]{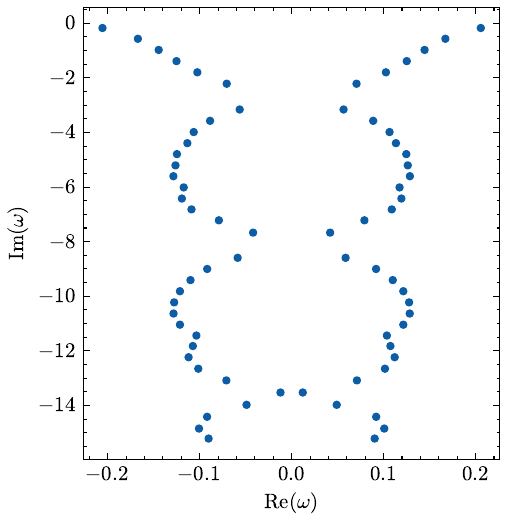}
         \caption{$N=1000$}
         \label{}
     \end{subfigure}
     \begin{subfigure}[b]{0.3\textwidth}
         \centering
         \includegraphics[width=\textwidth]{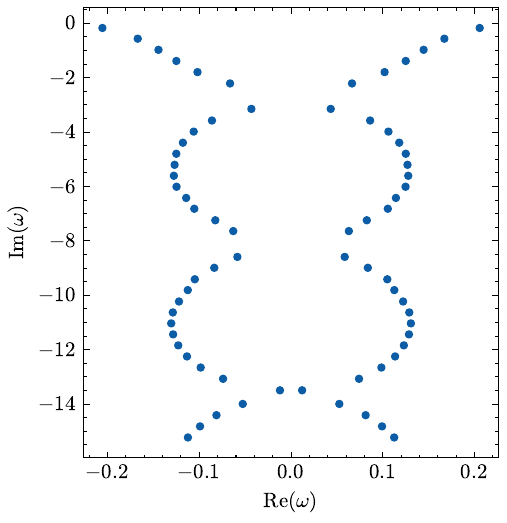}
         \caption{$N=1600$}
         \label{}
     \end{subfigure}
     \caption{Evolution of the QNM spectrum with the number of terms $N$ in the continued fraction for \textit{(above)} the BH given by equation \eqref{metric} for $a_0=0.05, \ P=0$ and for \textit{(below)} the BH studied in \cite{Moreira_2023}.}
     \label{itcomp}
\end{figure}

In the spectrum \ref{big_a0} one can see that in addition of the branch crossing there are some oscillations appearing in the highly damped part of the spectrum. Note that this figure was obtained with the maximum precision possible for our setup (\texttt{Mathematica} software on a laptop). The question is then to know whether those oscillations are a physical feature of this BH's spectrum (as it was found for another effective LQG BH in \cite{Moreira_2023}), or if they are just a characteristic of a lack of precision. 

To answer this question, we computed the QNM frequencies for $a_0=0.05$ and $P=0$ and we varied the precision by changing the number of terms $N$ in the continued fraction. The results, using the 15-order recursion satisfying by the coefficients of the expansion \eqref{ansatz} of $\Psi(r)$, are displayed in figure \ref{itcomp}. One notices that oscillations are apparent for $N=200$ but disappear when $N$ is increased. Doing the same process for the ABBV BH studied in \cite{Moreira_2023} show that the big oscillations stay consistent when increasing the precision while the small oscillations disappear when $N$ is taken bigger. Those big oscillations are therefore physically relevant and are a feature of this model. On the other side we conclude that the small oscillations appearing in the spectrum of \ref{big_a0} only express a lack of precision.

\section{Correspondence with circular null geodesics in the eikonal limit}
\label{eikonal_appendix}

\subsection{WKB approximations and analytical QNM formula}

Looking at the formula for the scalar effective potential \eqref{pot}, it can be expressed in the eikonal limit ($l\rightarrow\infty$) as:
\be 
V(r)\sim \lp l+\frac{1}{2}\rp^2 \frac{f(r)}{h(r)}.
\ee
The first order WKB calculation gives the quantization rules for the QNM:
\be
\omega^2= V_0-i\lp n+\frac{1}{2}\rp\sqrt{-2\frac{\dd^2 V_0}{\dd x^2}}
\,,\qquad
\omega
\underset{V_0\gg1}\approx \sqrt{V_0}-i\lp n +\frac{1}{2}\rp \sqrt{-\frac{1}{2V_0}\frac{\dd^2 V_0}{\dd x^2}},
\ee
where the subscript 0 means that the quantities were evaluated at the radius $r=r_0$ for which the effective potential is maximum. The maxima condition translates to:
\be
f_0h_0'=f_0'h_0.
\label{cond0}
\ee
Computing the second derivative of the effective potential V with respect to the tortoise coordinate $x$ defined in \eqref{eq:tortoise} and using the condition \eqref{cond0} leads to:
\be 
\frac{\dd^2 V_0}{\dd x^2}= \sqrt{\frac{f_0g_0}{h^2_0}(f_0h_0''-h_0f_0'')}.
\ee
Finally we get a simple formula giving the QNM frequencies in the eikonal limit:
\be
\omega \approx \lp l+\frac{1}{2}\rp \sqrt{\frac{f_0}{h_0}} -i \lp n+\frac{1}{2}\rp \sqrt{\frac{g_0}{2h_0}(f_0h_0''-h_0f_0'')}
\label{eikonalQNM0}
\ee
\\
Now let us look at circular null geodesics and show that the QNM frequency in the eikonal limit can be solely be expressed in terms of parameters of the circular null geodesics. First worked out for Schwarzschild black holes, this correspondence  was extended to a subclass of stationary and spherically symmetric black holes in \cite{Cardoso_2009}. Let us review their approach.

Let us first have a look at the geodesics of our spherically symmetric space-time \eqref{metric}. Thanks to the spherical symmetry, we can focus on equatorial orbits and consider the Lagrangian
\be 
\cL = \frac{1}{2}\lc f(r)\Dot{t}^2-\frac{1}{g(r)}\Dot{r}^2-h(r)\Dot{\phi}^2\rc.
\ee
$\phi$ denotes the angular coordinate.
The generalized momenta read:
\begin{align}
p_t=f(r)\Dot{t} \equiv E
\,,\qquad
p_{\phi}=-h(r)\Dot{\phi} \equiv -L
\,,\qquad
p_r= -\frac{1}{g(r)}\Dot{r}\,,
\end{align}
where $E$ refers to the energy and $L$ to the angular momentum.
Due to the Killing symmetries under time and angular translations, both $E$ and $L$ are conserved.
The Hamiltonian thus reads:
\begin{align}
    \cH
    = p_t\Dot{t}+p_r\Dot{r}+p_{\phi}\Dot{\phi}-\cL
    = E\Dot{t}-\frac{1}{g(r)}\Dot{r}^2-L\Dot{\phi}.
\end{align}
This Hamiltonian vanishes for null geodesics. Then, defining the effective potential as $V_r\equiv \Dot{r}^2$, we get:
\be 
V_r=g(r)\lc \frac{E^2}{f(r)}-\frac{L^2}{h(r)}\rc.
\ee 
The circular null geodesics (for $r=r_c$) follow the conditions $V_r=V_r'=0$, thus leading to:
\be
\label{condc}
\frac{E}{L}=\pm\sqrt{\frac{f_c}{h_c}}
\,,\qquad\textrm{and}\qquad
f_ch_c'=h_cf_c'
\,,
\ee
where the subscript $c$ means that the quantities are evaluated at the radius $r_c$ of a circular null geodesic. Notice the equivalence of conditions \eqref{condc} and \eqref{cond0}. In fact, $r_0$ (for which the effective potential \eqref{pot} of the Schrödinger-like equation is maximal) coincides with the radius of the null circular geodesic $r_c$: $r_0=r_c$.
The coordinate angular velocity for $r=r_c$ is 
\be
\Omega_c = \frac{\Dot{\phi}}{\Dot{t}}=\sqrt{\frac{f_c}{h_c}}.
\ee 
On the other hand the Lyapunov exponent can be computed via \cite{Cardoso_2009}:
\be
\lambda=\sqrt{\frac{V_r''(r_c)}{2\Dot{t}^2}}\,,
\qquad\textrm{with}\quad
V_r''(r_c)=\frac{g_cL^2}{f_ch_c^2}\lc f_ch_c''-h_cf_c''\rc\,,
\ee 
leading to the formula 
\be 
\lambda=\sqrt{\frac{g_c}{2h_c}\lc f_ch_c''-h_cf_c''\rc}\,.
\ee
Notice the agreement with the WKB formula \eqref{eikonalQNM0}, which can now be written directly in terms of $\Omega$ and $\lambda$ as
\be
\omega \approx \lp l+\frac{1}{2}\rp \Omega_c -\lp n+\frac{1}{2}\rp \lambda\,.
\ee
We conclude that QNM frequencies in the eikonal limit of any metric of the form \eqref{metric} are given in terms of the angular velocity $\Omega_c$ and the Lyapunov exponent $\lambda$, giving the instability timescale of the geodesic.

\subsection{QNMs in the Eikonal Limit: Numerics}
\label{app:numerics-eikonal}

We compared the QNM frequencies obtained via Leaver's method and those resulting from the analytical prediction given above in equation \eqref{eikonalQNM0}. Here are the numerics for the six first modes  at angular momentum $l=100$ and values of the quantum parameters $a_0\in[0,0.2]$ and $P\in[0,0.002]$.

\begin{table}[h!]
    \centering
    \begin{tabular}{|c||c|c|c|}
    \hline
    & Continued fraction & Circular null geodesics\\
    \hline
    $n=0$ & missing & 38.6821 - 0.192448$i$\\
    $n=1$ & 38.6815 - 0.577359$i$& 38.6821 - 0.577344$i$\\
    $n=2$ & 38.6794 - 0.962288$i$& 38.6821 - 0.962241$i$\\
    $n=3$ & 38.6762 - 1.34725$i$& 38.6821 - 1.34714$i$\\
    $n=4$ & 38.672 - 1.73226$i$& 38.6821 - 1.73203$i$\\
    $n=5$ & 38.6666 - 2.11734$i$& 38.6821 - 2.11693$i$\\
    \hline
    \end{tabular}
    \caption{$l=100 \ (a_0=0, \ P=0)$}
\end{table}

\begin{table}[h!]
    \centering
    \begin{tabular}{|c||c|c|c|}
    \hline
    & Continued fraction & Circular null geodesics\\
    \hline
    $n=0$ & missing & 38.6066 - 0.192326$i$\\
    $n=1$ & 38.6056 - 0.576986$i$& 38.6066 - 0.576977$i$\\
    $n=2$ & 38.6035 - 0.961665$i$& 38.6066 - 0.961628$i$\\
    $n=3$ & 38.6002 - 1.34638$i$& 38.6066 - 1.34628$i$\\
    $n=4$ & 38.5959 - 1.73114$i$& 38.6066 - 1.73093$i$\\
    $n=5$ & 38.5905 - 2.11596$i$& 38.6066 - 2.1155$i$\\
    \hline
    \end{tabular}
    \caption{$l=100 \ (a_0=0.1, \ P=0)$}
\end{table}

\begin{table}[h!]
    \centering
    \begin{tabular}{|c||c|c|c|}
    \hline
    & Continued fraction & Circular null geodesics\\
    \hline
    $n=0$ & 38.3855 - 0.191977$i$& 38.382 - 0.19193$i$\\
    $n=1$ & 38.3844 - 0.575938$i$& 38.382 - 0.57579$i$\\
    $n=2$ & 38.3821 - 0.959918$i$& 38.382 - 0.95965$i$\\
    $n=3$ & 38.3787 - 1.34393$i$& 38.382 - 1.34351$i$\\
    $n=4$ & 38.3742 - 1.72799$i$& 38.382 - 1.72737$i$\\
    $n=5$ & 38.3686 - 2.11211$i$& 38.382 - 2.11123$i$\\
    \hline
    \end{tabular}
    \caption{$l=100 \ (a_0=0.2, \ P=0)$}
\end{table}

\begin{table}[h!]
    \centering
    \begin{tabular}{|c||c|c|}
    \hline
    & Continued fraction & Circular null geodesics\\
    \hline
    $n=0$ & missing  & 38.6821 - 0.192448$i$\\
    $n=1$ & 38.6815 - 0.577359$i$ & 38.6821 - 0.577344$i$\\
    $n=2$ & 38.6794 - 0.962288$i$ & 38.6821 - 0.962241$i$\\
    $n=3$ & 38.6762 - 1.34725$i$ & 38.6821 - 1.34714$i$\\
    $n=4$ & 38.672 - 1.73226$i$ & 38.6821 - 1.73203$i$\\
    $n=5$ & 38.6666 - 2.11734$i$ & 38.6821 - 2.11693$i$\\
    \hline
    \end{tabular}
    \caption{$l=100 \ (a_0=0, \ P=0)$}
\end{table}

\begin{table}[h!]
    \centering
    \begin{tabular}{|c||c|c|}
    \hline
    & Continued fraction& Circular null geodesics\\
    \hline
    $n=0$ & 38.7858 - 0.192879$i$& 38.7857 - 0.192879$i$\\
    $n=1$ & 38.7848 - 0.578643$i$& 38.7857 - 0.578634$i$\\
    $n=2$ & 38.7826 - 0.964427$i$& 38.7857 - 0.964389$i$\\
    $n=3$ & 38.7794 - 1.35025$i$&  38.7857 - 1.35015$i$\\
    $n=4$ & 38.7752 - 1.73611$i$& 38.7857 - 1.7359$i$\\
    $n=5$ & 38.7699 - 2.12204$i$& 38.7857 - 2.12166$i$\\
    \hline
    \end{tabular}
    \caption{$l=100 \ (a_0=0, \ P=0.001)$}
\end{table}

\begin{table}[h!]
    \centering
    \begin{tabular}{|c||c|c|}
    \hline
    & Continued fraction &  Circular null geodesics\\
    \hline
    $n=0$ & 38.8892 - 0.193307$i$& 38.8891 - 0.193306$i$\\
    $n=1$ & 38.8882 - 0.579927$i$& 38.8891 - 0.579918$i$\\
    $n=2$ &  38.8860 - 0.966567$i$& 38.8891 - 0.966529$i$\\
    $n=3$ & 38.8828 - 1.35324$i$ & 38.8891 - 1.35314$i$\\
    $n=4$ & 38.8786 - 1.73997$i$ & 38.8891 - 1.73975$i$ \\
    $n=5$ & 38.8732 - 2.12675$i$ & 38.8891 - 2.12636$i$\\
    \hline
    \end{tabular}
    \caption{$l=100 \ (a_0=0, \ P=0.002)$}
\end{table}

\section{Comparison with existing papers}
\label{app:others}

\subsection{Other numerical QNM computation for this metric}

In this section we compare our results with those given by other papers. To this end, we use Leaver's method to compute the QNM frequencies for a set of parameters values which were examined in papers \cite{Momennia:2022tug} and \cite{Liu:2020ola}. 

\begin{table}[!h]
    \centering
    \begin{tabular}{|c||c|c|c|}
     \hline
     $P$ & $\omega_{01}$ & $\omega_{02}$ & $\omega_{12}$ \\
     \hline
     0.0 &  $0.616510 - 0.203845 i$ & $0.967288 - 0.193518 i$ &  $0.927701 - 0.591208 i$ \\
     \hline
     0.1 & $0.747795 - 0.238366 i$  &  $1.24123 - 0.236866 i$ &  $1.1975 - 0.722273 i$ \\ 
     \hline
     0.2 &  $0.930412 - 0.281022 i$ & $1.5514 - 0.279967 i$ &  $1.50566 - 0.851966 i$ \\
     \hline
     0.3 & $1.13063 - 0.320604 i$ &  $1.89259 - 0.320076 i$ & $1.84721 - 0.971978 i$ \\
     \hline
     0.4 & $1.34331 - 0.354122 i$ &  $2.25607 - 0.354137 i$ & $2.21337 - 1.07317 i$ \\ \hline \midrule
      \multicolumn{4}{c}{ }  \\ \hline \midrule
     $P$ & $\omega_{01}$ & $\omega_{02}$ & $\omega_{12}$ \\
     \hline
     0.0 &  (4.89\%)(4.89\%)& (0.0122\%)(0.0122\%) & (0.00921\%)(0.00921\%) \\
     \hline
     0.1 & (0.00438\%)(4.62\%)  & (0.00574\%)(2.71\%) & (0.0116\%)(7.84\%) \\ 
     \hline
     0.2 &  (0.00258\%)(9.07\%) & (0.00209\%)(5.35\%)  &  (0.00399\%)(15.7\%) \\
     \hline
     0.3 & (0.00258\%)(13.1\%)  & (0.00399\%)(7.81\%) & (0.00116\%)(23.1\%)  \\
     \hline
     0.4 & (0.0109\%)(16.8\%)  & (0.00412\%)(9.99\%) & (0.00172\%)(29.8\%)  \\
     \hline
    \end{tabular}
    \caption{QNM frequencies $\omega_{nl}$ for $a_0=0.01$ and different values of $P$ (computed with Leaver's method) (\textit{up}), along with the relative difference with results respectively given in table I of \cite{Momennia:2022tug} and table I of \cite{Liu:2020ola} (calculated using sixth order WKB) (\textit{down}).}
    \label{compP}
\end{table}

First we look at QNM frequencies with $a_0=0.01$ and several values of $P$. Results are given in table \ref{compP}. Both \cite{Momennia:2022tug} and \cite{Liu:2020ola} computed the QNM frequencies using the sixth order WKB formula, but \cite{Momennia:2022tug} already stressed some discrepancies with the results of \cite{Liu:2020ola} and pointed a misuse of the WKB technique. Our results are in very good agreement with \cite{Momennia:2022tug} results, apart for the mode $\omega_{01}$ which can be explained by the limitations of the WKB method.

\begin{table}[!h]
    \centering
    \begin{tabular}{|c||c|c|c|}
    \hline
    $a_0$ & $\omega_{00}$ & $\omega_{01}$ & $\omega_{02}$\\
    \hline
    0 & $0.220910 - 0.209791 i$\ (0.00442\%) & $0.585872 - 0.19532 i$\ (0.00557\%)  & $0.967288 - 0.193518 i$\ (0.00219\%)  \\
    \hline
    0.25 & $0.215543 - 0.208251 i$\ (0.0218\%) & $0.577896 - 0.194472 i$\ (0.00464\%) & $0.955228 - 0.192759 i$\ (0.00509\%)  \\
    \hline
    0.5 & $0.200814 - 0.205718 i$\ (0.00793\%) & $0.55764 - 0.192996 i$\ (0.00681\%) & $0.924517 - 0.19127 i$\ (0.00365\%)  \\
    \hline
    \end{tabular}
    \caption{QNM frequencies $\omega_{nl}$ for $P=0$ and different values of $a_0$ (computed with Leaver's method), along with the relative difference with results given in table III of \cite{Momennia:2022tug} (calculated using asymptotic iteration method).}
    \label{compa0}
\end{table}
Then we look at QNM frequencies with $P=0$ and several values of $a_0$ and compare our results with those of \cite{Momennia:2022tug} which were computed using the asymptotic iteration method. Results are displayed in table \ref{compa0} and are again in very good agreement. 

\subsection{Comparison with another effective LQG  BH}

\begin{figure}
\centering
\includegraphics[width=0.35\textwidth]{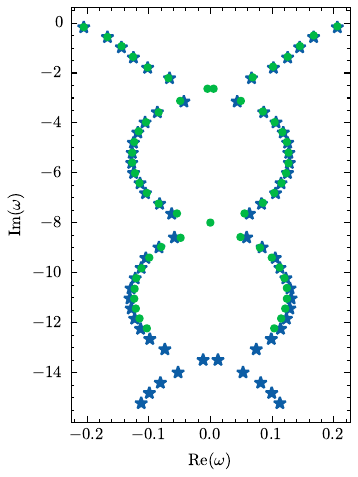}
\caption{\centering Comparison with the results of the paper \cite{Moreira_2023} for the ABBV BH with $r_0=0.4$. The green dots are the QNM frequencies extracted from \cite{Moreira_2023}. The blue stars are the QNM frequencies computed with the same code which was used for the results of the present paper.}
\label{r0comp}
\end{figure}

Recently the QNM of another effective LQG  BH -- the ABBV BH, see \eqref{catalogue} -- were investigated in \cite{Moreira_2023} using the continued fraction method. The metric of this Loop Quantum BH is parameterised with a single parameter $r_0$ -- $0<r_0<1$ -- defining the radius of the minimal spacelike hypersurface separating the BH interior from the White Hole area. The paper \cite{Moreira_2023} found that branches oscillations start appearing from some value of $r_0$. We applied our code to this BH metric and computed the corresponding QNM frequencies in order to compare them to the results shown in \cite{Moreira_2023}. The spectra are displayed in figure \eqref{r0comp}.

Our results match very well with with those of \cite{Moreira_2023} even if some small discrepancies start emerging for frequencies with high imaginary parts.

\section{Comparison with other methods}
\label{app:compare}

\subsection{Third order WKB approximation}

The WKB method is well known in quantum mechanics where it is used to study solutions of the Schrödinger equation. It has been first applied to the problem of black holes scattering by Schutz and Will in 1985 \cite{Schutz:1985km}. This semi-analytic technique can be used for any barrier type effective potential having constant values at the boundaries. The idea is to match the WKB solution at the event horizon with the one at infinity with a Taylor series expansion. The third order approximation reads \cite{Iyer:1986nq}:
\be
\omega^2\approx V_0 + \sqrt{-2V_0''}\Lambda - i\lp n+\frac{1}{2}\rp \sqrt{-2V_0''}(1+\Omega),
\ee
where the coefficients $\Lambda$ and $\Omega$ are given in terms of the potential by
\beq
\Lambda &=&
\frac{1}{\sqrt{-2V_0''}}
\left[
\frac{1}{8}\lp \frac{V_0^{(4)}}{V_0''}\rp \lp \frac{1}{4}+\alpha^2\rp-\frac{1}{288}\lp\frac{V_0'''}{V_0''}\rp^2\lp 7+60\alpha^2)\rp
\right]
\,,\\
\Omega &=&
\frac{1}{-2V_0''}
\Bigg[
\frac{5}{6912}\lp\frac{V_0'''}{V_0''}\rp^4\lp 77+188\alpha^2\rp
-\frac{1}{384}\lp \frac{V_0'''^2 V_0^{(4)}}{V_0''^3}\rp \lp51+100\alpha^2\rp
\\
&&+\frac{1}{2304}\lp \frac{V_0^{(4)}}{V_0''}\rp^2\lp 67 +68 \alpha^2\rp +\frac{1}{288}\lp \frac{V_0'' V_0^{(5)}}{V_0''^2}\rp\lp 19+28\alpha^2\rp - \frac{1}{288}\lp\frac{V_0^{(6)}}{V_0''}\rp \lp 5+4\alpha^2\rp \Bigg]\,. \nn
\eeq
$\alpha$ is given by $\alpha = n+\frac{1}{2}$. The derivatives are taken with respect to the tortoise coordinate and the subscript $0$ indicates that the potential and its derivatives are evaluated at the maximum of the potential.\\
We use this technique to compute some QNM frequencies and compare them to those computed thanks to the continued fraction method. The results are given in table \ref{WKBcomp}. As expected, we see that the WKB approximation is not accurate for small $l$ and $n\sim l$, yet there is a good correspondence for $l=10$ and even for $l=2, \ n=0$.

\begin{table}[!h]
    \centering
    \begin{tblr}{|c|c||c|c|}
    \hline
    $l$ & $n$ & Leaver's method & WKB approximation \\
    \hline
    \hline
    1 & 0 & $0.664098 - 0.216737i$  &  $0.624548 + 0.035814i$ \\
    \hline
    2 & {0 \\1} & { $1.09948 - 0.215064i$ \\  $1.05758 - 0.656426i$ } & { $1.08868 - 0.157007i$ \\ $0.870372 - 0.270846i$ } \\
    \hline
    10 & {0 \\ 1 \\  2} & { $4.60194 - 0.214128i$ \\ $4.59122 - 0.643049i$ \\ $4.56996 - 1.07396i$ } & { $4.60181 - 0.211615i$ \\ $4.5888 - 0.625707i$ \\ $4.55635 - 1.01356i$ } \\
    \hline
    \end{tblr}
    \caption{Comparison between some QNM frequencies with $a_0=0.01\ \& \ P=0.05$ computed using Leaver's method and third order WKB approximation.}
    \label{WKBcomp}
\end{table}

\subsection{Spectral decomposition}

We compare our numerical results for QNM values with another fully numerical computation, developed in~\cite{Jansen:2017oag}, which relies on a decomposition of the mode function $\Psi$ of~\eqref{schro} onto Chebychev polynomials. This decomposition transforms the QNM computation problem into a generalized eigenvalue problem. It is available as a Mathematica package dubbed \texttt{QNMSpectral}.

In table~\ref{tab:compar-QNMSpectral}, we compare the values obtained with Leaver's method and with the \texttt{QNMSpectral} package. We observe that they match up to numerical precision (see App.~\ref{convergence}). The spectral decomposition method has a precision that decreases with the overtone number $n$: therefore, it can only be trusted for the computation of the first overtones, and cannot yield as many QNM values as Leaver's method.

\begin{table}[!htb]
    \centering
    \begin{tblr}{|c|c||c|c|}
    \hline
    $l$ & $n$ & Leaver's method & \texttt{QNMSpectral} \\
    \hline
    \hline
    1 & {0 \\ 1} & {$0.6640980607007495 - 0.2167366809416785 i$ \\ $0.6035617699094376 - 0.6782460547001679 i$}  & { $0.66409806 - 0.21673668i$ \\ $0.6036 - 0.6782 i$ } \\
    \hline
    2 & {0 \\1} & { $1.0994814189825621 - 0.21506432221876454i$ \\  $1.057577864185476 - 0.6564262637378766i$ } & { $1.099481419-0.2150643222i$ \\ $1.0575779-0.6564263i$ } \\
    \hline
    10 & {0 \\ 1 \\  2 \\ 3 \\ 4 \\ 5} & { $4.601941156497556 - 0.21412772420261542i$ \\ $4.591221597721874 - 0.6430494149913577i$ \\ $4.5699640505758765 - 1.073960723533072i$ \\ $4.538534312241122 - 1.5081560527130353i$ \\ $4.497486021962071 - 1.9468763641684066i$ \\ $4.447561990485726 - 2.39127816179962i$} & { $4.601941156-0.2141277242i$ \\ $4.591221598-0.643049415i$ \\ $4.5699641-1.0739607i$ \\ $4.538534-1.508156i$ \\ $4.4975 - 1.9469i$ \\ $4.448-2.391i$} \\
    \hline
    \end{tblr}
    \caption{Comparison between some QNM frequencies with $a_0=0.01\ \& \ P=0.05$ computed using Leaver's method and the \texttt{QNMSpectral} Mathematica package.}
    \label{tab:compar-QNMSpectral}
\end{table}

\section{Convergence of the continued fraction method}
\label{convergence}

\begin{figure}[!h]
    \centering
    \includegraphics{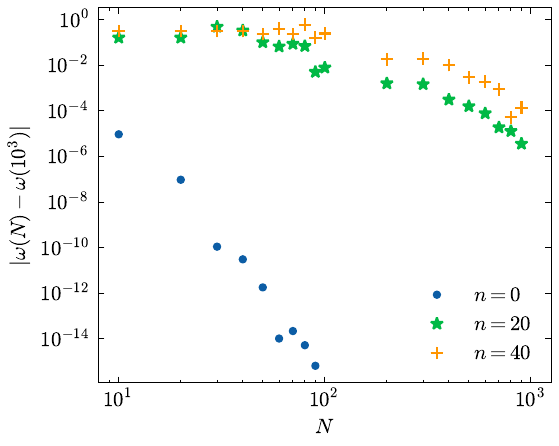}
    \caption{Convergence of QNM frequencies with respect to the number of terms in the continued fraction. $(a_0=0.01,P=0.05,l=2)$}
    \label{convergence_plot}
\end{figure}

It is important to ensure that the computed frequencies converge towards a fixed value when the number of terms in the continued fraction \eqref{eq:contfrac-schwa} $N$ is increased. We show on figure \ref{convergence_plot} the evolution of the precision of the QNM values with respect to $N$ by computing the relative difference with the value computed at a high precision (here $N=10^{3}$). Without any surprise, we see that the precision is much better for the fundamental mode (notice that the threshold was set at $10^{-16}$ and that this precision is reached from $N=100$ for $\omega_0$). The precision then decreases for higher overtones. In particular, one can note that a precision of $10^{-6}$ can be reached with $N=900$ for $\omega_{20}$.

As a matter of information, we noticed that the convergence is better when considering $P=0$, which allowed us to compute more QNM frequencies in the $a_0$-spectrum \ref{big_a0} while still ensuring a good precision.

\bibliographystyle{bib-style}
\bibliography{biblio.bib}

\end{document}